\documentclass{aa}  

\usepackage{graphicx}
\usepackage{txfonts}

\newcommand{\eff}{_{\textrm{\tiny eff}}} 
\renewcommand{\sun}{_{\odot}} 
\newcommand{\MS}{\textrm{MS}} 
\newcommand{\SD}{{\textrm{SD}}} 

\begin{document} 

   \title{Alone but not lonely: Observational evidence that binary interaction is always required to form hot subdwarf stars}
   \titlerunning{Binary interaction is always required to form hot subdwarf stars}

  \author{Ingrid Pelisoli\inst{1}\thanks{pelisoli@astro.physik.uni-potsdam.de}, Joris Vos\inst{1}, Stephan Geier\inst{1}, Veronika Schaffenroth\inst{1}, Andrzej S. Baran\inst{2}}
  \authorrunning{Pelisoli et al.}

   \institute{
   Institut f\"{u}r Physik und Astronomie, Universit\"{a}t Potsdam, Haus 28, Karl-Liebknecht-Str. 24/25, D-14476 Potsdam-Golm, Germany
   \and
   ARDASTELLA Research Group, Institute of Physics, Pedagogical University of Krakow, ul. Podchor\c{a}\.zych 2, 30-084 Krak\'ow, Poland
   }

   \date{Received MONTH DD, YYYY; accepted MONTH DD, YYYY}

 
  \abstract
   {Hot subdwarfs are core-helium burning stars that show lower masses and higher temperatures than canonical horizontal branch stars. They are believed to be formed when a red giant suffers an extreme mass-loss episode. Binary interaction is suggested to be the main formation channel, but the high fraction of apparently single hot subdwarfs (up to 30\%) has prompted single star formation scenarios to be proposed.
   }
   {We investigate the possibility that hot subdwarfs could form without interaction by studying wide binary systems. If single formation scenarios were possible, there should be hot subdwarfs in wide binaries that have undergone no interaction.
   }
   {Angular momentum accretion during interaction is predicted to cause the hot subdwarf companion to spin up to the critical velocity. The effect of this should still be observable given the timescales of the hot subdwarf phase. To study the rotation rates of companions, we have analysed light curves from the Transiting Exoplanet Survey Satellite (TESS) for all known hot subdwarfs showing composite spectral energy distributions indicating the presence of a main sequence wide binary companion. If formation without interaction were possible, that would also imply the existence of hot subdwarfs in very wide binaries that are not predicted to interact. To identify such systems, we have searched for common proper motion companions with projected orbital distances of up to 0.1~pc to all known spectroscopically confirmed hot subdwarfs using {\it Gaia} DR2 astrometry.}
   {We find that the companions in composite hot subdwarfs show short rotation periods when compared to field main sequence stars. They display a triangular-shaped distribution with a peak around 2.5 days, similar to what is observed for young open clusters. We also report a shortage of hot subdwarfs with candidate common proper motion companions. We identify only 16 candidates after probing 2938 hot subdwarfs with good astrometry. Out of those, at least six seem to be hierarchical triple systems, in which the hot subdwarf is part of an inner binary.}
   {The observed distribution of rotation rates for the companions in known wide hot subdwarf binaries provides evidence of previous interaction causing spin-up. Additionally, there is a shortage of hot subdwarfs in common proper motion pairs, considering the frequency of such systems among progenitors. These results suggest that binary interaction is always required for the formation of hot subdwarfs.}


   \maketitle
%

\section{Introduction}

Hot subdwarf stars are underluminous objects that lie in the extreme horizontal branch \citep[EHB,][]{heber1986}. They have thinner hydrogen envelopes ($M_\textrm{env} < 0.02~M\sun$) than canonical horizontal branch stars, being thus unable to sustain hydrogen shell burning. Because of their thin envelopes, they also appear hotter than their canonical counterparts. They show temperatures $T\eff > 20\,000$~K and surface gravities in the range $4.5 < \log~g < 6.5$ \citep[for a complete review of hot subdwarfs, see][]{heber2016}.

The characterisation of hot subdwarfs is of interest for many fields of astronomy, from extragalactic to stellar astrophysics. In extragalactic astrophysics, they are known to account for a large fraction of the ultraviolet excess observed in early-type galaxies \citep{oconnell1999}. They contribute to Galactic astrophysics by allowing us to probe the gravitational potential of the Milky Way, and in particular the mass of the dark matter halo \citep{tillich2011}, because they can be among the fastest stars in the Galaxy \citep[e.g. the hypervelocity star US708 in][]{geier2015}. They can also show pulsations, being remarkable targets for asteroseismology and allowing us to place constraints in stellar evolution theory \citep{charpinet2009}. Hot subdwarfs in close binaries can qualify as type Ia supernova progenitors \citep[e.g.][]{maxted2000} and/or verification sources for future space-based gravitational wave detectors such as the Laser Interferometer Space Antenna \citep[{\it LISA}, see][]{kupfer2018}. 

Despite their ubiquitous importance in astronomy, their formation remains puzzling. The consensus is that hot subdwarfs are the progeny of low- to intermediate-mass stars that have undergone an episode of enhanced mass-loss at the tip of the red-giant branch (RGB). Binary interaction is the main evoked explanation for such an episode.

Three main binary evolution scenarios have been described in detail by \citet{han2002, han2003}: (i) common envelope evolution, (ii) stable Roche-lobe overflow (RLOF), and (iii) the merger of two He white dwarfs. In scenario (i), the more massive star in the binary would evolve to the RGB and
fill its Roche lobe, inefficiently transferring mass to the companion and leading to the formation of a common envelope; this envelope is eventually ejected, leaving an exposed He-burning core in a close binary (orbital period of up to a few days). If the mass transfer is instead stable (scenario ii), the outer layers of the hot subdwarf progenitor are slowly stripped away by its companion. This channel leads to the formation of hot subdwarfs with main sequence companions in wide binaries (orbital periods of tens to hundreds of days).

Supporting these two channels, a large fraction of hot subdwarfs are found in binaries. About one-third are found in close binaries with periods from hours to a few days, consistent with scenario (i), mostly showing white dwarf or M-type main sequence stars as companions \citep{maxted2001, morales2003, geier2011, copperwheat2011}, whereas 30--40\% show composite-colours and/or spectra indicating the presence of K to F type companions in wide orbits \citep{stark2003}. The latter are often referred to as composite hot subdwarfs \citep{vos2018}. 

It should be noted that the large observed fraction of composite binaries is not necessarily an indication of a previous mass-transfer phase forming the hot subdwarf. Given evolutionary timescales and the age of the Universe, hot subdwarf progenitors descend from main sequence stars of type F or earlier, which have binary fractions $\gtrsim 50$\% \citep[e.g.][]{duchene2013}. Therefore, a similar fraction of hot subdwarfs in composite binaries could be expected, even if interaction was not required to form them.

A first attempt to verify whether the visible companions in composite hot subdwarf systems were sufficiently close to support previous interaction was made by \citet{heber2002}, who tried to resolve 19 composite systems with the Hubble Space Telescope. Considering the observed separation distribution of progenitors in binary systems, about one third of the observed sample should have been resolved. However, only two systems were resolved, one of which turned out be a hierarchical triple system with the sdB being part of a close, single-lined binary.

Although this result already indicated a significant deviation of the separation distribution of composite hot subdwarfs from their progenitor systems, the stable RLOF-channel could only be finally proven when the first orbital solutions of composite sdB systems were determined \citep{deca2012, barlow2012, barlow2013, vos2012, vos2013}. The derived long periods also required an update of the binary evolution models to be consistent with observations \citep{chen2013}. A dedicated survey of a small sample of composite systems bright enough to be observed with high-resolution spectrographs showed that a high fraction of systems shows radial velocity variability and, importantly, high values of $v \sin i$ ($> 10$~km/s) for the companions, suggesting a high rate of previous interactions \citep{vos2018}.

On the other hand, for a large fraction of hot subdwarfs, of up to 30\%, no companions have been found. This fraction is even much higher for the hot subdwarfs found in globular clusters \citep[see][and references therein]{latour2018}. In the binary framework, these single objects are mainly explained as the result of the remaining scenario: a merger of two He white dwarfs. In fact a few single hot subdwarfs are found to be fast rotators, supporting this theory \citep[e.g. EC~22081-1916 and SB290 in][respectively]{geier2011b, geier2013}. However, most of the single hot subdwarfs have been found to be very slow rotators both from analyses of the rotational line broadening \citep{geier2012} and from asteroseismic analyses of space-based light curves \citep{baran2009, baran2012, pablo2012, reed2014}. In addition, the agreement between model predictions and observations is still poor \citep{zhang2009}, and the predicted broad mass distribution of systems resulting from mergers seems to be at odds with the observed narrow distribution \citep{schneider_david_2019_3428841}. Moreover, the companions of helium white dwarfs are found to be mostly canonical to massive white dwarfs \citep{brown2020}, suggesting a shortage of progenitors for the merger scenario.

Alternative formation scenarios relying on single star evolution have been proposed. \citet{dcruz1996} proposed that strong stellar wind mass loss in the RGB phase could place some objects in the EHB. \citet{sweigart1997} suggested that helium mixing from the hydrogen shell into the envelope, driven by internal rotation, could cause enhanced mass-loss in the RGB. However, if these single evolution scenarios were indeed possible, there must also be hot subdwarfs in wide binaries that have undergone no interaction. In these systems, neither of the components should show any measurable traces of previous interactions, such as increased rotation rates due to transfer of angular momentum observed in the confirmed RLOF systems \citep{vos2018} or pollution due to accreted matter. In particular, given the properties of the progenitor systems \citep[e.g.][]{derosa2014, moe2017, elbadry2019}, there must be a significant fraction of very wide, resolved visual binaries observable as common proper motion pairs.

In this work, we investigate the possibility that hot subdwarfs could form without binary interaction by (i) characterising rotation rates for the companions in known composite binaries believed to have formed via the RLOF channel and (ii) searching for common proper motion companions to all spectroscopically confirmed hot subdwarfs. In Section \ref{TESS}, we analyse light curves from the Transiting Exoplanet Survey Satellite (TESS) for known hot subdwarfs in composite binaries in search for evidence of mass transfer. Mass transfer causes angular momentum to be gained by the companion stars, which are thus predicted to spin up \citep[e.g.][]{kippenhahn1977, popham1991}. In Section \ref{gaia}, we perform a search for common proper motion companions to hot subdwarfs with projected orbital distances up to 20\,000~AU ($\approx0.1$~pc), where no interaction is expected. We present a discussion of our results in Section~\ref{discussion}, and conclude in Section~\ref{conclusion}.


\section{TESS light curves for composite hot subdwarfs}
\label{TESS}

The TESS mission \citep{ricker2015} was launched in 2018 to obtain high-precision photometry from space with the goal of finding nearby rocky exoplanets. The nominal two-year mission observed 26 sky sectors, each with a field of $24^{\circ} \times 90^{\circ}$, for 27 days. The sectors cover over 90\% of the sky, avoiding only a narrow band around
the ecliptic already partially explored by the K2 mission \citep{howell2014}. There is overlap between different sectors, therefore the total coverage can be much larger than 27~days (up to 351~days for stars around both ecliptic poles). TESS obtained images of each sector every 2~seconds, which were used for guiding. These 2~second images were stacked into 20~seconds, 2~minutes, or 30~minute cadence images that could be downloaded to the ground. The 20~second cadence postage-stamps were only downloaded for a small number of very bright asteroseismology targets, whereas 2~minute postage-stamps were obtained for a large number of objects proposed by the community. Finally, every pixel observed by TESS in the nominal two-year mission was downloaded at 30 minute cadence.

Although TESS has been designed as an exoplanet mission, the cadence and high precision are also of particular interest for studying binary stars and intrinsic stellar variability. For evolved compact stars such as hot subdwarfs and also white dwarfs, the TESS Asteroseismic Science Consortium (TASC) Working Group 8 (WG8) has proposed an extensive variability survey including all known evolved compact stars brighter than 16th magnitude. Up to Sector 20, 1125 hot subdwarf and candidates from the catalogues of \citet{geier2017} and \citet{geier2019} were observed in 2~minute cadence. 

The catalogue of spectroscopically confirmed hot subdwarfs from \citet{geier2017} has recently been updated taking into account input from the {\it Gaia} data release 2 \citep[DR2,][]{gaiadr2}, as well as the latest releases of the Large Sky Area Multi-Object Fibre Spectroscopic Telescope \cite[LAMOST,][]{lamost}. We have crossmatched the updated catalogue \citep{geier2020} with the list of observed stars from the TASC WG8. We identified 156 stars classified as composite hot subdwarfs with main sequence companions (sdO/B + MS, + A, + F, + G, + K) observed up to Sector 20. We note that, although cooler companions are predicted, these systems would not be detected as composite because the hot subdwarf completely dominates the light, and therefore M companions are not included in our analysis. Similarly, in systems with earlier-type companions, the hot subdwarf would not be detected in optical observations, therefore such systems are also not included in our sample. We note, however, that mass transfer from hot subdwarfs has been invoked to explain the fast rotation of Be stars \citep{wang2018}, therefore there is already evidence for mass transfer in binary systems of hot subdwarfs with early-type companions.

\begin{figure}
\centering
\includegraphics[width=\hsize]{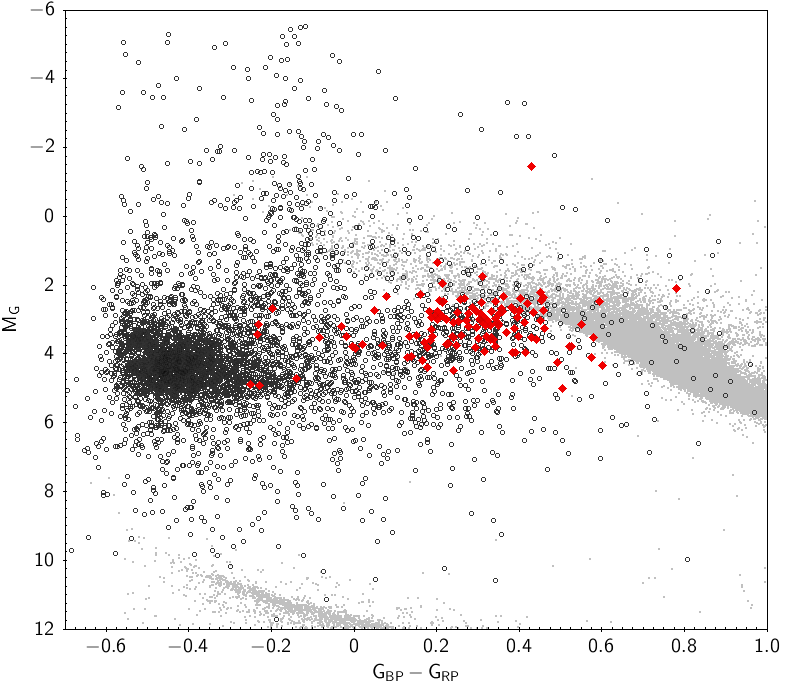}
\caption{Colour-magnitude diagram showing all stars from the catalogue of \citet{geier2020} as dark grey circles. The composite systems analysed in this work are shown as red diamonds. At this point, no quality cuts have been performed in the astrometry, which explains the high spread. Sample C of \citet{lindegren2018} is shown in light grey to facilitate the location of the main sequence and other evolutionary stages.}
\label{GaiaHR}
\end{figure}

To remove objects that might not be real binaries, but rather have been misclassified due to contamination by a nearby star, we excluded objects with another source in {\it Gaia} DR2 within 5 arcsec, unless they were confirmed radial velocity variables. We also excluded objects classified as close binaries in the literature, since those are formed by the common envelope and not the RLOF channel, as well as objects whose fit to the spectral energy distribution (SED) indicated they were actually single stars. We were then left with 123 composite hot subdwarfs observed with TESS (Fig.~\ref{GaiaHR}).

We have searched for variability in the light curves of all 123 composite objects using a Lomb-Scargle periodogram \citep{lomb1976, scargle1982}. We have used the light curves made available by the TESS Science Processing Operations Center (SPOC), and specifically the PDCSAP flux, which corrects the simple aperture photometry (SAP) to remove instrumental trends, as well as contributions to the aperture expected to come from neighbouring stars other than the target of interest given a pre-search data conditioning (PDC). This is particularly relevant for TESS because the pixel size is nearly 21". The pipeline also provides an estimate of how much of the flux in the aperture belongs to the target systems in the {\tt CROWDSAP} parameter. To avoid possible zero-point inconsistencies between different sectors, the reported fluxes were divided by the mean flux in each sector for each star. We have also performed sigma-clipping to exclude any measurements more than 5-sigma away from the median value.

Our initial analysis consisted of calculating the periodogram up to the Nyquist frequency, with sampling to give ten points per significant periodogram peak, phase-folding the data to the dominant peak, and inspecting the periodogram and phase-folded light curve to confirm any variability. We have found 90 out of the 123 stars (73\%) to show periodic variability. The false-alarm probability (FAP) of the dominant period for objects classified as variable was in most cases FAP $\lesssim 10^{-20}$. For two objects, the periodogram is typical of $g$-mode hot subdwarf pulsators \citep{green2003}, as shown in Fig.~\ref{pulsators}. Both of these are new discoveries; they are discussed in Appendix~\ref{sec:puls}, and the identified periods are listed in Table~\ref{table:puls}.

\begin{figure}
\centering
\includegraphics[width=\hsize]{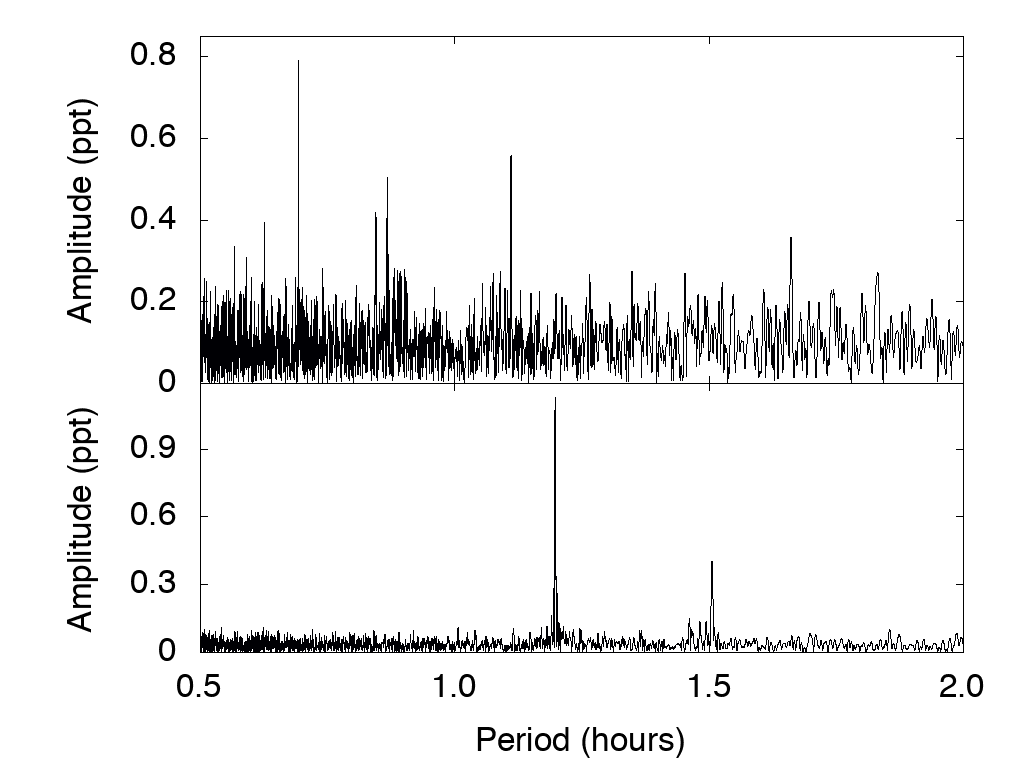}
\caption{TIC~71013467 (top) and TIC~158235404 (bottom), two new pulsators that are part of our sample. They show periods in the range of 45~min to 2~hours, typical of $g$-mode hot subdwarf pulsators \citep{green2003}.}
\label{pulsators}
\end{figure}

For the remaining objects, the variability cannot be attributed to pulsations of the hot subdwarf star. It instead likely originates in the main sequence companion. One possible scenario is an interplay between stellar activity and rotation, as seen for many Kepler \citep[e.g.][]{reinhold2015} and K2 objects \citep[e.g.][]{reinhold2020}. The activity causes temperature and therefore brightness changes across the surface of the star, which are seen as periodic variability as the star rotates. To test whether that was the cause of the variability, we have analysed the light curves for these 90 variable objects following the approach of \citet{reinhold2013}, as summarised below.

Firstly, a Lomb-Scargle periodogram was calculated up to the Nyquist frequency. We have set the minimal frequency to $2/T$, where $T$ is the time span of the light curve (e.g. $T=27$~days for objects observed in only one sector). This assumes that a minimum of two cycles are necessary to detect the variability. As above, we used an oversampling factor of ten to determine the frequency step, that is $\delta f = 1/(10T)$. Once the dominant peak was identified, we have performed pre-whitening to identify a second possible peak. Computing a Lomb-Scargle periodogram is equivalent to fitting a sine wave to the data, hence the pre-whitening consists on simply subtracting the fitted sine wave from the data, and computing the periodogram of the residuals. As we are only interested in the dominant period and not in all observed periods, this process was performed only once.

The main goal of the pre-whitening was to verify that the dominant period was not an alias. In some cases spots can be located on opposite sides of the star, in which case the dominant peak will actually correspond to half the rotation period. Following the prescription of \citet{reinhold2013}, we have compared the two periods with largest power in the periodogram and, if the difference between twice the first period and the second period was less than 5\%, we selected the longer one, which is more likely the correct period.

\begin{figure*}
\centering
\includegraphics[width=\hsize]{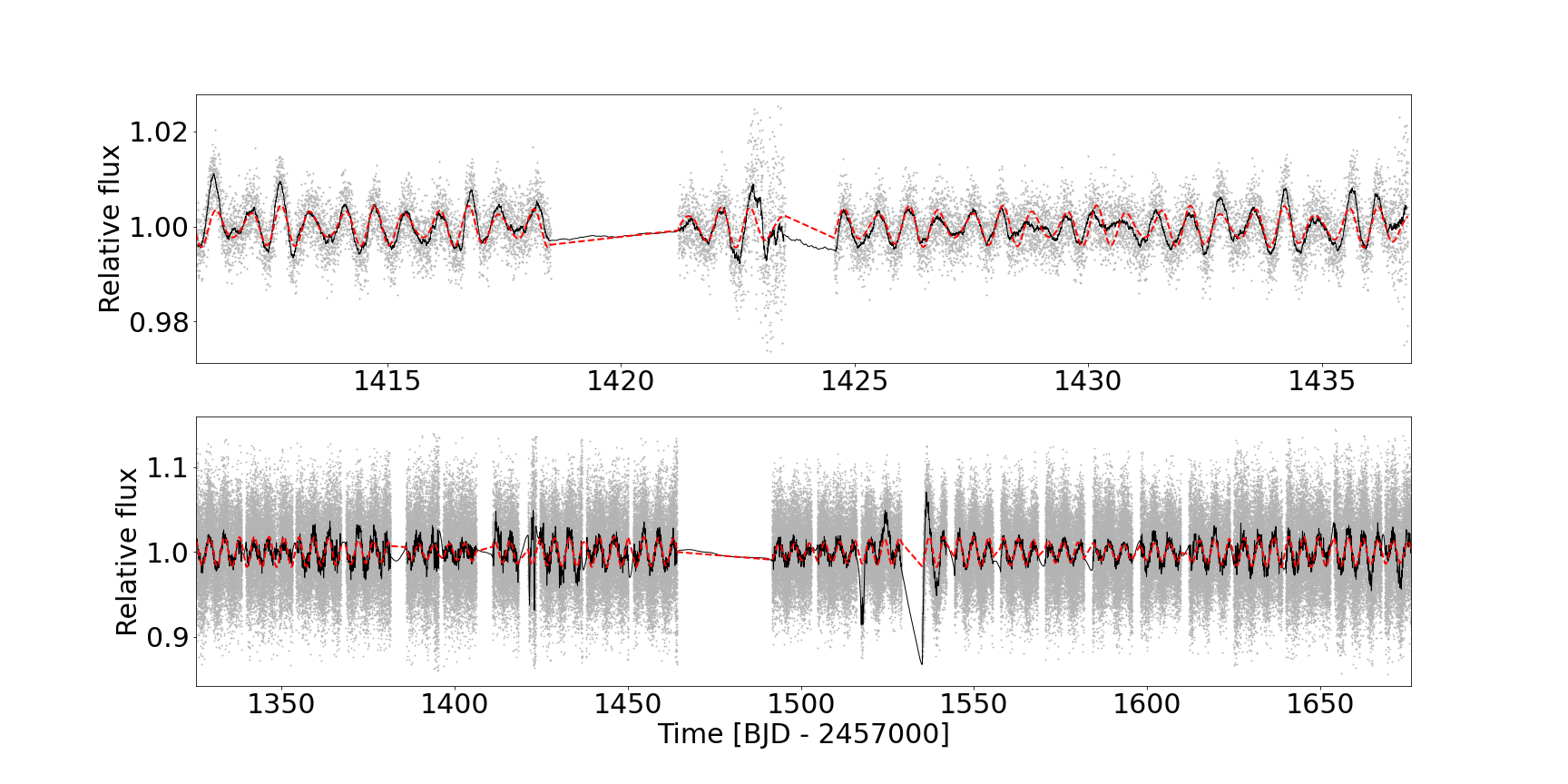}
\caption{TIC~12528447 (top) and TIC~382518318 (bottom), two systems for which the variability has been attributed to rotation of the main sequence companion. The former was observed in only one sector, while the latter is in the TESS continuous viewing zone and has 2~min cadence data for 12 sectors. The grey dots are the TESS data, the black line is a running mean every 50 points, and the dashed red line is the multi-component fit.}
\label{rotvar}
\end{figure*}

For many stars, differential rotation is present, in which case a second dominant peak which is not an alias is detected. To identify this, we checked whether the period $P_2$ was within 30\% of $P_1$ \citep[which suggests it is consistent with differential rotation,][]{reinhold2013}. Defining $\alpha = |P_2 - P_1| / \textrm{max}(P_1, P_2)$, we verified if $P_2$ was such that $\delta f < \alpha < 0.3 $, where the lower limit accounts for the frequency resolution of each light curve of $1/(10T)$. If $\alpha$ was within these limits, $P_2$ was accepted as a second significant period arising from differential rotation, otherwise it was discarded. This step was performed to improve accuracy on $P_1$, since the second sine wave with period $P_2$ can have a significant effect on the light curve. Last, sine parameters for both $P_1$ and $P_2$, when significant, were used as input for a global sine fit, summing both sine waves and allowing for both periods to vary. We show two examples in Fig.~\ref{rotvar}.

Finally, as our goal is to study a clear sample of rotators, we attempted to remove from the sample objects whose cause of variability is possibly not rotation or whose periods are not well determined. Systems were flagged by visual inspection when the light curve was not obviously periodic, or when the periods seemed to be only marginally significant. We also compared the number of zero crossings with the prediction for a sine wave. A single sine wave has two zero crossings per period, therefore the observed period is indeed due to rotation we should expect a number of crossings of the order of $N = T\times 2/P_1$. A number of crossings higher than that hints at a different cause for variability, such as stellar pulsations or irregular variations. We estimated the number of zero crossings by smoothing the data using a boxcar average with a width of $5 \times P_1$, followed by an average every ten points to reduce fluctuations due to uncertainty in flux. The number of crossings was not used as a hard-limit, given that more crossings can be shown if, for example, there are spots on opposite sides of the star, or if the rotation pattern exhibits double dips \citep[as seen predominantly for slow rotators,][]{Basri2018}.

\begin{figure*}
\centering
\includegraphics[width=\hsize]{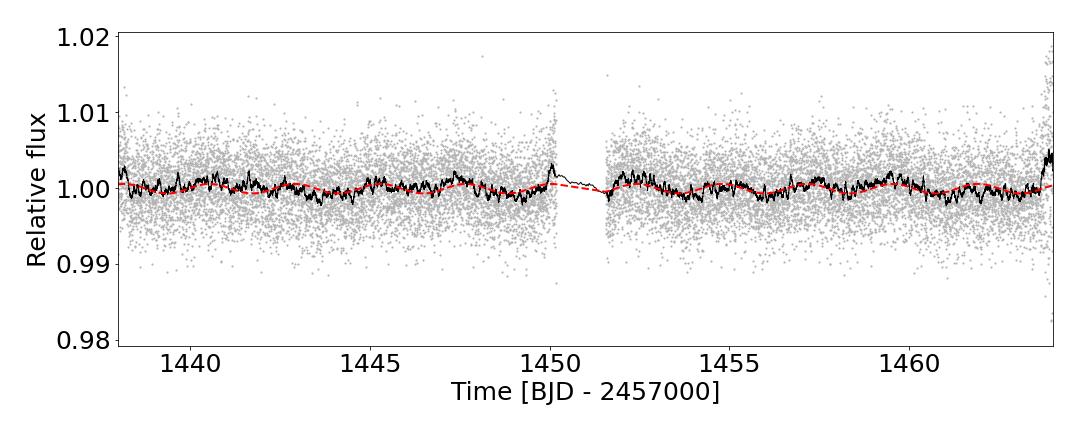}
\caption{TIC~13069774, a variable star showing a number of crossings more than twice the predicted value for sinusoidal variation.}
\label{othervar}
\end{figure*}

\begin{figure}
\centering
\includegraphics[width=\hsize]{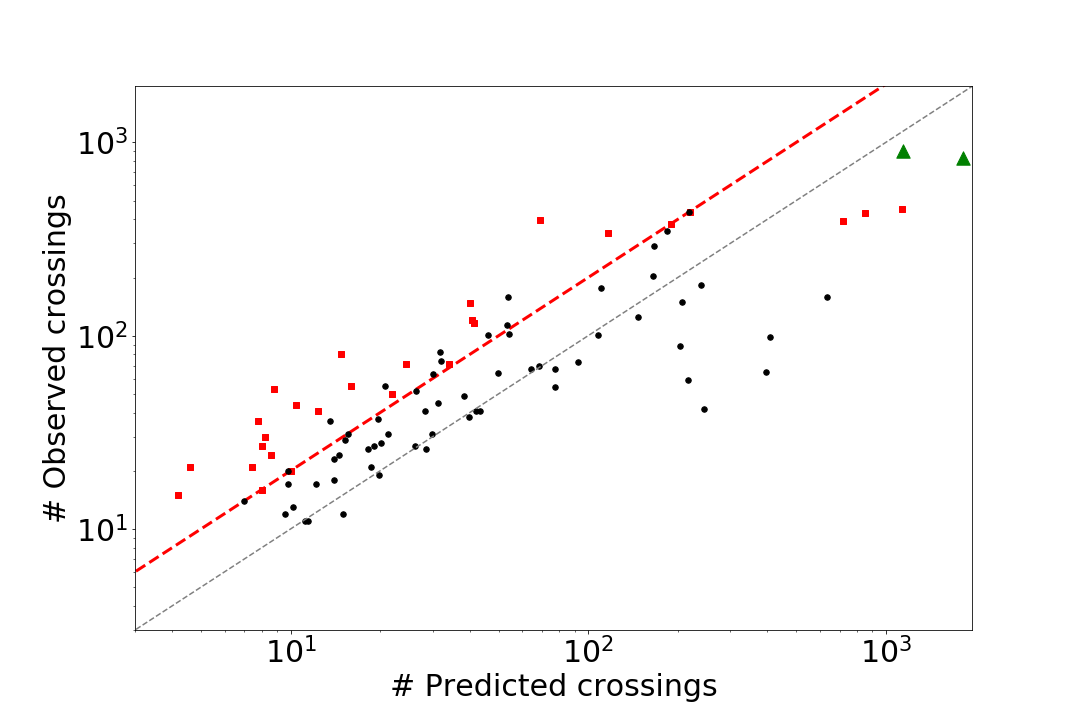}
\caption{Observed number of zero crossings as a function of the predicted number of crossings for all 90 variables. The two $g$-mode hot subdwarf pulsators are shown as green triangles. The two dashed lines represent equality (bottom, in grey) and observed number of crossings equal to twice the number of predicted crossings (top, in red). Objects not included as rotational variables are shown as red squares. Objects with a high number of predicted crossings due to showing short periods often show less observed crossings because TESS data is not continuous, presenting a gap every $\simeq 13$ days when data was being transmitted to Earth.}
\label{crossings}
\end{figure}

We instead further inspected objects showing more than two crossings per period. Many of those had already been flagged as uncertain during our initial inspection. An example is shown in Fig.~\ref{othervar}. Other objects not previously flagged showed higher number of crossings mainly because of fluctuations due to the uncertainty in flux. We further flagged as uncertain three objects showing a number of crossings lower than predicted that clustered closely to the two identified pulsators (see Fig.~\ref{crossings}). We were then left with 61 composite systems whose variability we interpret as due to rotation, listed in Table~\ref{rot}. The remaining 29 variable objects are listed in Table~\ref{table:other}. An interpretation on the origin of their variability is out of the scope of this work. It is quite possible that they are also rotators, but we do not treat them as such so that our sample consists only of stars whose variability can be safely interpreted as due to rotation. Objects not observed to show periodic variability (NOV) are listed in Table~\ref{table:NOV} with the respective detection limits.

\begin{figure}
\centering
\includegraphics[width=\hsize]{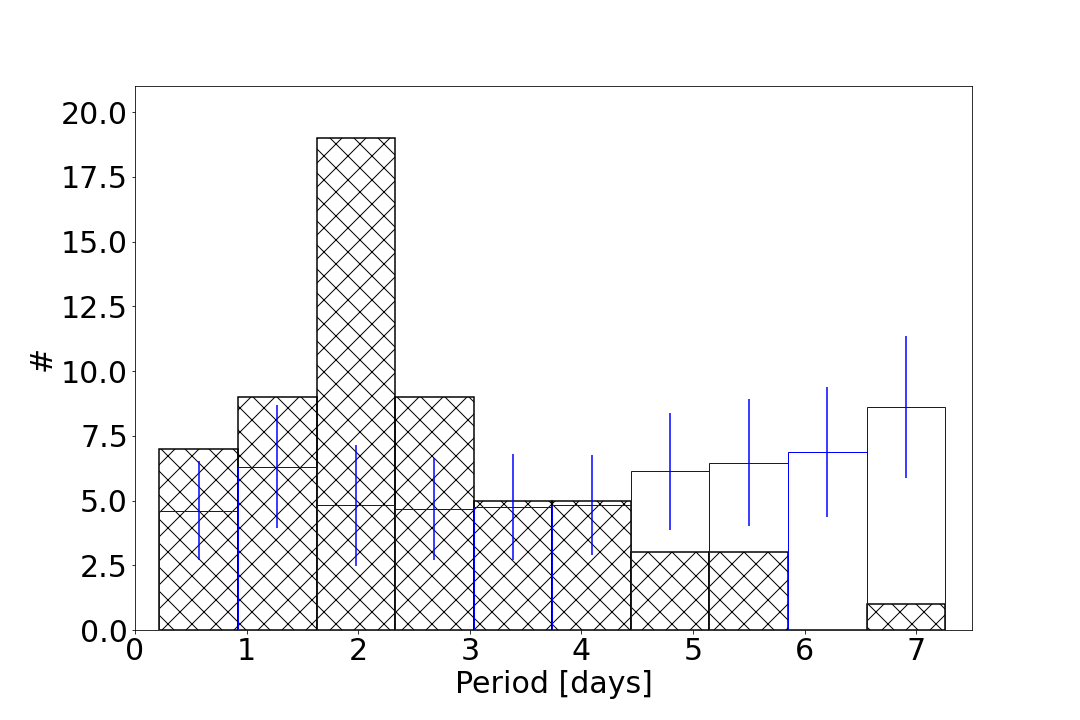}
\caption{Number histogram of the obtained rotational periods for the hot subdwarf companions (black, hatched). As a comparison, we show a number histogram obtained from drawing a sample of objects from \citet{reinhold2015} with the same temperature distribution as our sample and periods within the detectable range (in blue). This was repeated a hundred times; the height of the histogram is the average over all realisations, whereas the error bar is the standard deviation.}
\label{periods}
\end{figure}

Figure~\ref{periods} shows, in black, the histogram of the obtained periods for the 61 presumably rotational variables. The mean period is 2.5~days, with a one-sigma spread of 1.4~days. To obtain a comparison with field main sequence stars with no hot subdwarf companions, we have used the data from \citet{reinhold2015}, who determined the rotational periods for more than 18\,500 stars observed by the Kepler mission. To account for the temperature distribution of our sample, which excludes cool companions that cannot be detected in composite binaries, we selected for each object in our sample the stars in \citet{reinhold2015} with same effective temperature (within 5\%).

The temperature for the main sequence stars in the systems in our sample was determined with a multi-component fit to the spectral energy distribution (SED), taking into account both the contribution of the hot subdwarf and of the companion. The SED fitting procedure uses literature photometry from APASS \citep{henden2015}, 2MASS \citep{skrutskie2006}, WISE \citep{cutri2012}, and {\it Gaia} \citep{gaiadr2, riello2018, evans2018}, and is constrained by the reddening obtained from the dust maps of \citet{lallement2019} and the {\it Gaia} parallax. To model the SED, T\"ubingen NLTE Model-Atmosphere package models \citep{werner2003} and Kurucz atmosphere models \citep{kurucz1979} are used for respectively the hot subdwarf star and the cool companion. A more detailed description of the SED fitting procedure is given in \citet{vos2017, vos2018b}.

Next, given a sample of stars with consistent temperature for each of our objects, we accounted for the fact that there is an upper limit to detectable periods by randomly selecting the period of one of the stars in the comparison sample considering only periods within our detectable range (we assumed that periods shorter than the maximum period in our sample, of $\simeq 7.5$~days, could be detected). We repeated this a hundred times, obtaining thus one period histogram for each realisation. We then calculated the mean histogram, as well as an uncertainty given by the standard deviation. This histogram is shown in blue in Figure~\ref{periods}.

The period distribution for main sequence companions to hot subdwarfs is clearly different from the distribution of field main sequence stars. This is confirmed by a two-sample Anderson-Darling test, which suggests that the null hypothesis that these two distributions are sampled from the same population can be safely rejected ($p\textrm{-value} < 0.001$).

\begin{figure}
\centering
\includegraphics[width=\hsize]{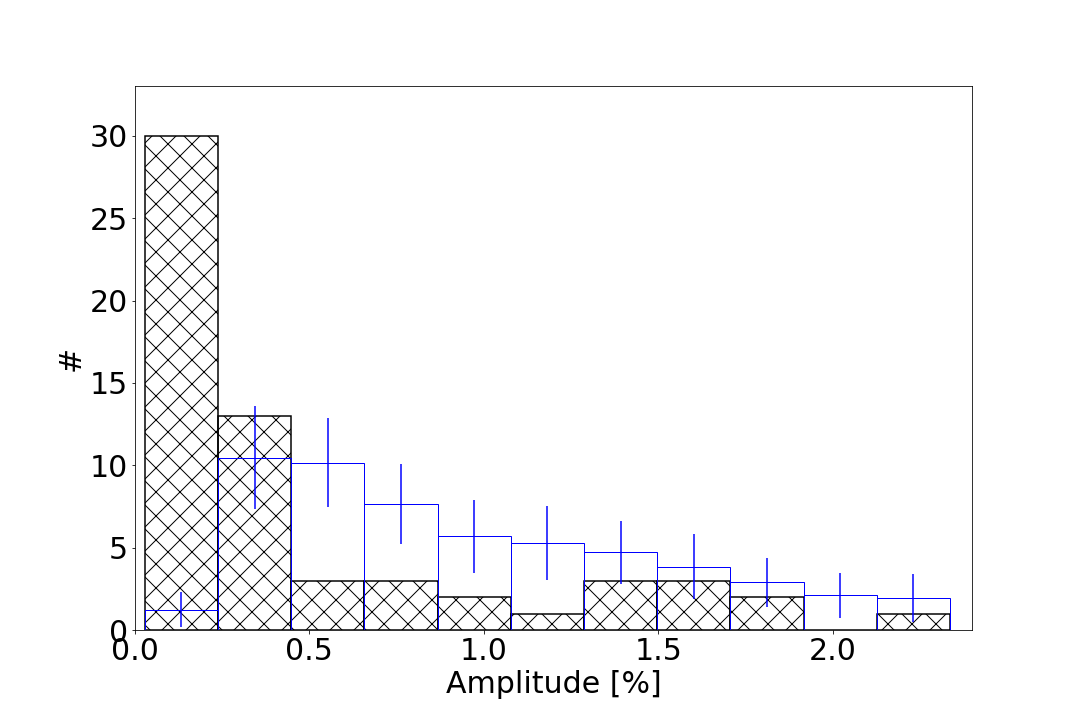}
\caption{Same as Figure~\ref{periods}, but for the detected amplitude of variability.}
\label{amplitudes}
\end{figure}

Performing a similar comparison for the amplitude of the observed photometric variability results in Figure~\ref{amplitudes}. To avoid differences due to variable precision in different magnitude ranges, in this case we have also constrained the comparison sample to have a similar magnitude distribution to our observed sample, by drawing comparison objects with magnitudes within 5\% of the value for each of our systems (or within 10\% when no pair was found). Albeit more subtle for the amplitude distribution compared to the period distribution, a two-sample Anderson-Darling test confirms that the behaviour of the companion stars also does not follow the trend of canonical main sequence stars ($p\textrm{-value} < 0.001$).

Figures \ref{periods} and \ref{amplitudes} suggest that main sequence stars with a subdwarf companion show, in average, shorter rotation periods and lower photometric variability amplitudes than the bulk of main sequence stars. The analysed main sequence stars should be as old as their hot subdwarf companions, which likely descend from F stars that have already evolved beyond the RGB, suggesting ages of at least a few billion years. The fact that their rotation is systematically faster than field main sequence stars, which show a large spread in age and thus rotation rates, can only be explained by accretion. As the hot subdwarf progenitor reaches the red giant phase and fills its Roche-lobe, the companion will accrete mass and be spun up to near the critical velocity, when the centrifugal acceleration exactly balances gravity. At this point, accretion can no longer cause the companion star to spin up, and it will start to spin down \citep{popham1991, paczynski1991, deschamps2013}. The short rotation periods can therefore be interpreted as evidence for previous interaction in these wide binaries. In fact, the shape of the observed period distribution as well as the observed range of periods resembles what is found for young open clusters \citep[e.g.][]{rebull2016, rebull2018}, in which stars are still rotating somewhat close to their critical velocities. A similar effect has also been observed for post-mass-transfer blue straggler stars \citep{leiner2018}.

The amplitudes are lower likely due to a combination of factors. On one hand, stellar activity is found to decrease with age \citep[e.g.][]{davenport2019}, therefore this difference in amplitude suggests that the composite companions are older than the main sequence stars in the comparison sample. However, metallicity also plays a role in the level of activity, and hence the lower amplitudes can also be suggesting that the companions have lower metallicities. There is no comprehensive study of the metallicity of main sequence stars in composite systems with hot subdwarfs, althouh a preliminary analysis of a few systems indicates that they seem to be slightly subsolar (Molina et al., {\it in preparation}).

\section{{\it Gaia} DR2 search for common proper motion companions}
\label{gaia}

One of the many applications of the unprecedented {\it Gaia} DR2 is the search for common proper motion pairs \citep[e.g.][]{elbadry2018, fouesneau2019}. Whereas close binaries ($\textrm{separations} \lesssim 20$~mas) are not resolved by {\it Gaia}, wider binaries can be identified as co-moving pairs thanks to the precise {\it Gaia} astrometry. For hot subdwarfs, radial velocity variability has been used as the main indicator for binarity, which limits the detection to objects with periods of a few tens of years at best \citep[the discovery of hot subdwarfs dates back to][although their evolutionary origin was only understood much later]{humason1947}. This is equivalent to separations smaller than $\simeq 20$~au, or $\simeq 20$~mas at a distance of 1~kpc, coinciding with objects unresolved by {\it Gaia}.

{\it Gaia} therefore opens a new window to study hot subdwarf binarity by allowing us, for the first time, to seek binary companions beyond 20~au. We have performed a search for common proper motion companions to spectroscopically confirmed hot subdwarfs in the catalogue of \citet{geier2020}. We followed the approach of \citet{fouesneau2019}, and searched for co-moving sources with proper motions consistent with those of the hot subdwarfs and candidates within 3~$\sigma$ according to:
\begin{equation}
    \frac{( (\mu_{\alpha})_{\SD} - (\mu_{\alpha})_{Gaia} )^2}{(\sigma_{\mu,\alpha})_{\SD}^2 + (\sigma_{\mu,\alpha})_{Gaia}^2} +
    \frac{( (\mu_{\delta})_{\SD} - (\mu_{\delta})_{Gaia} )^2}{(\sigma_{\mu,\delta})_{\SD}^2 + (\sigma_{\mu,\delta})_{Gaia}^2} \leq
    ( 3~\textrm{mas/yr} )^2,
\end{equation}
where $\mu_{\alpha}$ is the proper motion in right ascension ($\alpha$), $\mu_{\delta}$ is the proper motion in declination ($\delta$), $\sigma_{\mu,\alpha}$ is the uncertainty in $\mu_{\alpha}$, and $\sigma_{\mu,\delta}$ is the uncertainty in $\mu_{\delta}$. The subscripts `SD' and `$Gaia$' refer to the queried hot subdwarfs, and to objects in the {\it Gaia} catalogue other than the hot subdwarf. Similarly, we have also restricted the parallax ($\varpi$) difference between the two sources to 3~sigma:
\begin{equation}
    \frac{( \varpi_{\SD} - \varpi_{Gaia} )^2}{(\sigma_\varpi)_{\SD}^2 + (\sigma_\varpi)_{Gaia}^2} \leq
    ( 3~\textrm{mas} )^2.
\end{equation}
Finally, we have only considered {\it Gaia} sources whose projected separation was smaller than 20\,000~AU ($\sim 0.1$~pc), because pairs with larger separations are likely to be eventually disrupted by external gravitational perturbations \citep{retterer1982, weinberg1987}. This also helps limiting the contamination by chance alignments, which grows rapidly at large separations \citep{andrews2017}. The ADQL query applying these conditions can be found in Appendix~\ref{adql}.

The query returned 299 matches for 237 stars. We next applied quality filters to the astrometry of both hot subdwarf and candidate companion, by selecting only objects with parallax uncertainties smaller than 20\%, and applying the quality filters of \citet{lindegren2018}, that is
\begin{equation}
    1.0 + 0.015\,(G_{BP} - G_{RP})^2 < E < 1.3 + 0.06\, (G_{BP} - G_{RP})^2, \label{ast1}
\end{equation}
where $E$ is the $\verb!phot_bp_rp_excess_noise!$, the photometric excess factor obtained by comparing fluxes in the $G_{BP}$ and $G_{RP}$ passbands to the total $G$ flux \citep[see e.g.][]{evans2018}, and
\begin{equation}
    u < 1.2\, {\tt max}(1, \exp(-0.2(G - 19.5))), \label{ast2}
\end{equation}
where $$u = \sqrt{ \frac{\tt astrometric\_chi2\_al}{({\tt astrometric\_n\_good\_obs\_al} - 5)} },$$
with {\tt astrometric\_chi2\_al} being the value of the chi-square statistic of the astrometric solution, and {\tt astrometric\_n\_good\_obs\_al} being the number of good observations; both are given in the {\it Gaia} DR2 table.

\begin{figure}
\centering
\includegraphics[width=\hsize]{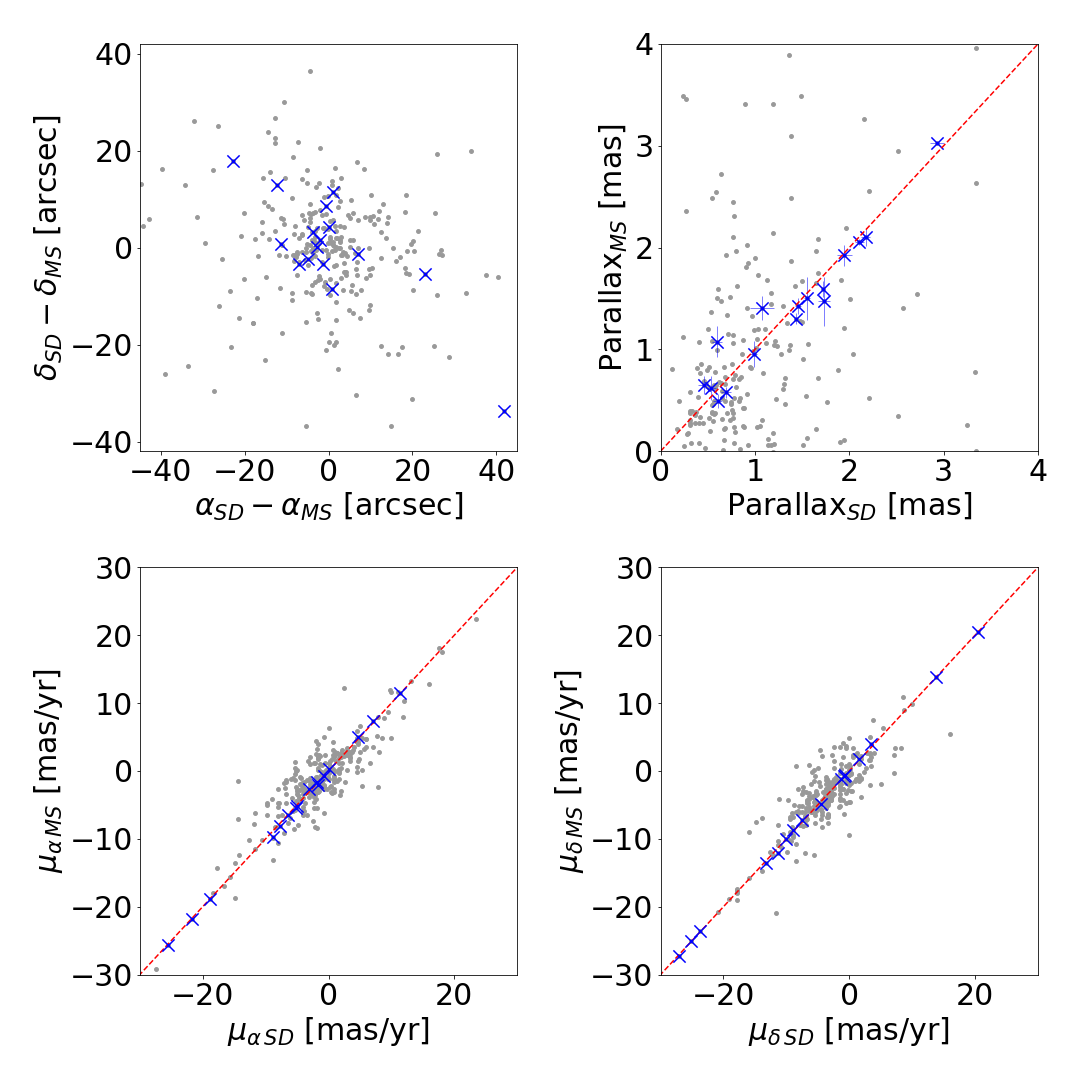}
\caption{Comparison between parameters for all matches (grey) and only those with good astrometry (blue). The top left panel shows the difference between coordinates of the hot subdwarf and the candidate companion. The top right panel compares the parallax of the two objects, whereas the bottom panels compare each component of the proper motion. The dashed lines represent equality.}
\label{parameters}
\end{figure}

\begin{figure}
\centering
\includegraphics[width=\hsize]{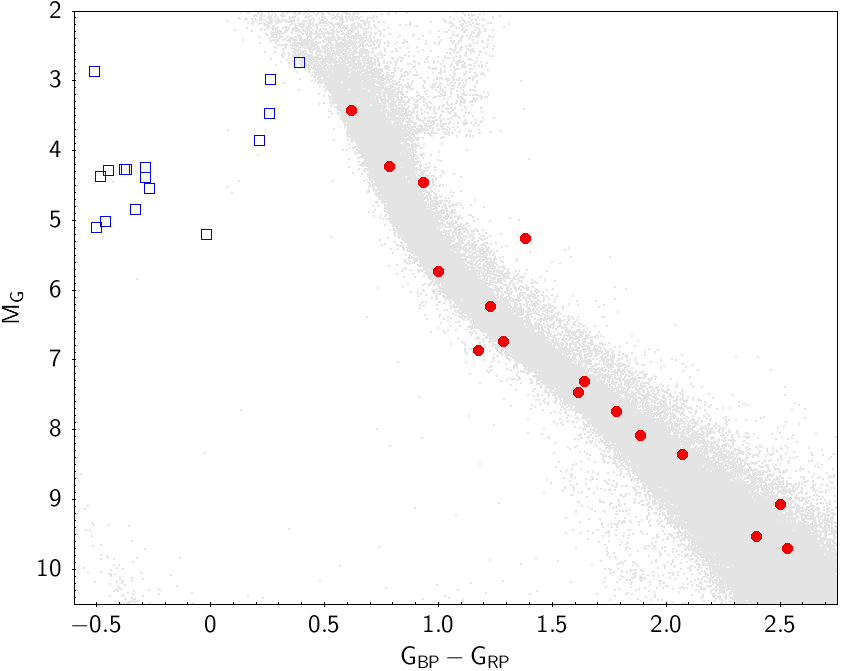}
\caption{Colour-magnitude diagram showing the position of all hot subdwarfs that have a common proper motion pair candidate. The hot subdwarfs are shown as open blue squares, and the candidate companions are shown as filled red circles. Light grey is the same as in Fig.~\ref{GaiaHR}.}
\label{cmd}
\end{figure}

This results on 16 common proper motion candidates, listed in Table~\ref{cppm}. The comparison between {\it Gaia} measurements for the pairs is shown in Fig.~\ref{parameters}. Figure~\ref{cmd} illustrates the position of both the hot subdwarfs and their candidate companions in a colour-magnitude diagram. We provide more details of each of these hot subdwarfs in Appendix~\ref{cpm_cand}.

Given the number of hot subdwarfs with good astrometry (Eqs. \ref{ast1} and \ref{ast2}), even if all our identifications are confirmed as hot subdwarfs with a common proper motion pair, this implies a fraction of only $\simeq 0.5$\%. We note that this fraction refers to the inhomogeneous sample of spectroscopically classified hot subdwarfs, and cannot be generalised given that no selection effects are taken into account. Yet, this very small incidence of common proper motion pairs among confirmed hot subdwarfs suggests that such systems are much less common than for their progenitors.

\section{Discussion}
\label{discussion}

We find that a large fraction of composite hot subdwarf binaries present evidence of interaction in the form of accretion, with at least 61 of the 123 analysed systems showing variability consistent with rotation of the companion. The rotation periods shown by the main sequence companions are significantly smaller than those shown by a comparison sample of main sequence stars.

The observed periods can be well described by a log-normal distribution with mean $\log P [\textrm{days}] = 0.35$ and standard deviation $\log P [\textrm{days}] = 0.27$. The period distribution for main sequence stars in \citet{reinhold2015} within the temperature range of the our sample can be described by a log-normal distribution with mean $\log P [\textrm{days}] = 1.19$ and standard deviation $\log P [\textrm{days}] = 0.26$. The distributions show therefore the same width, but the mean is shifted towards much lower values for the hot subdwarf companions. The longest period in our sample has about 10\% chance of coming from the distribution with larger mean, with all other periods showing probabilities lower than 5\%. However, we see no evidence for bi-modality in our sample, and a Shapiro-Wilk test cannot reject the null hypothesis that the logarithms of the observed periods follow a normal distribution (assuming a confidence level of 99\%), therefore we interpret all observed periods to come from the same distribution.

The observed amplitudes of the photometric variability are also found to be significantly lower than for the comparison sample. This reflects the fact that the hot subdwarf companions are an older, lower-metallicity population than the stars in the comparison sample. Although the observed stars are rejuvenated by accretion, which explains their short orbital periods, the accretion would not affect their interiors significantly \citep[given that accretion rates are fairly small,][]{toonen2012, vos2020}, hence their level of activity is unaffected and the amplitudes remain low. In other words, these stars appear young if their rotational periods are considered, but the amplitudes reveal them to be old.

\begin{figure}
\centering
\includegraphics[width=\hsize]{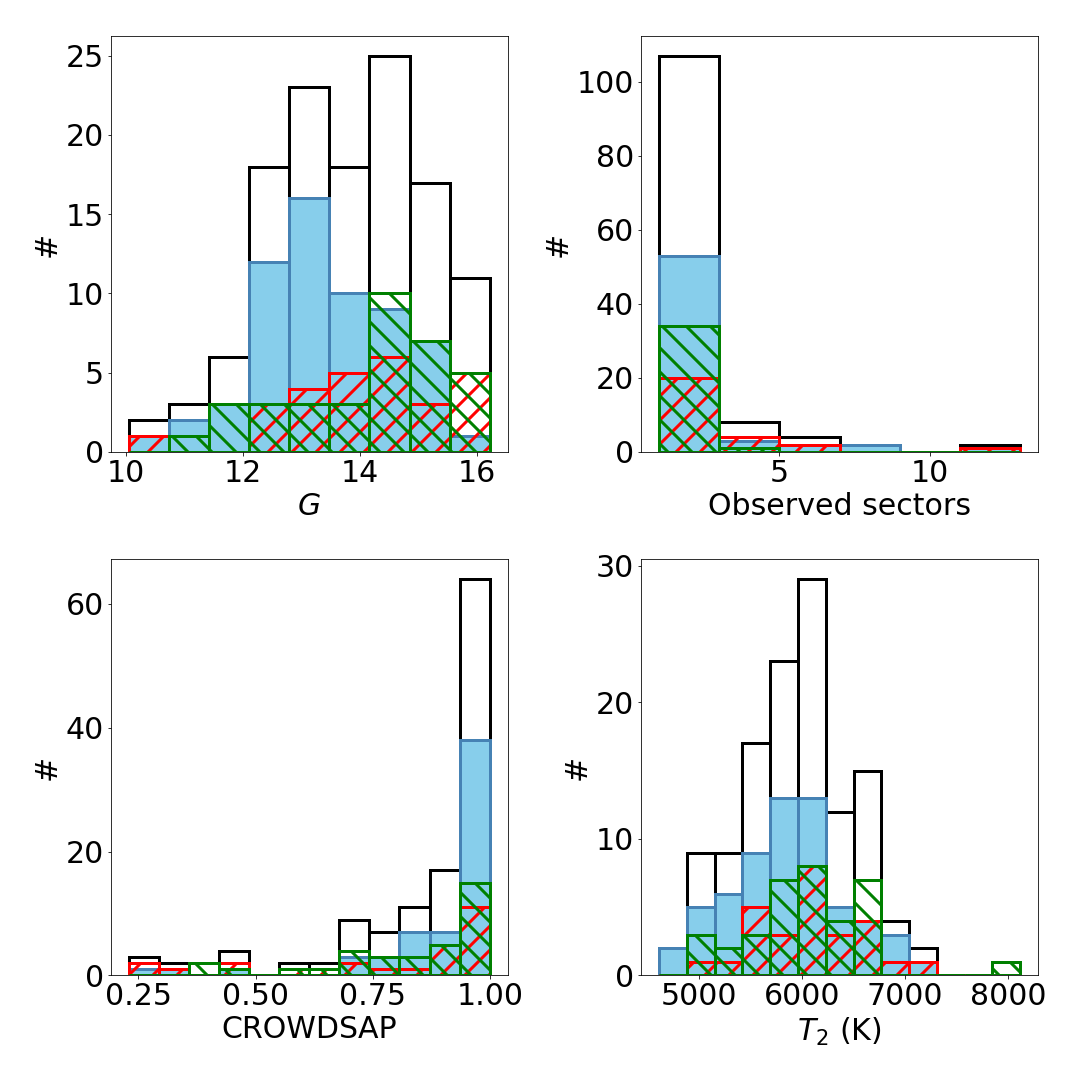}
\caption{Number histograms comparing the distribution of {\it Gaia} $G$ magnitude (top left), TESS observed number of sectors (top right) and {\tt CROWDSAP} parameter (bottom left), and companion effective temperature (bottom right) between the whole sample of 123 composites (unfilled, black), 61 rotational variables (filled, blue), 33 NOV systems (\textbackslash\textbackslash, green), and 29 other variable systems (//, red). As these 29 systems seen to follow the distribution of the entire sample, they are excluded from our statistical analysis.}
\label{hists}
\end{figure}

For a further 29 stars, variability is detected, but the periods are less significant, or the light curve is not obviously periodic. These objects seem to be an unremarkable subsample of the complete sample of 123 composite systems, with $G$ magnitude, {\tt CROWDSAP}, number of observed sectors, and effective temperature of the companion all similarly distributed to the entire sample, as shown in Fig.~\ref{hists}. This is also indicated by two-sample KS tests, in which the null hypothesis that the parameters for these 29 stars are drawn from the same population as the parameters for the whole sample of 123 objects cannot be rejected. In order to compile a sample of clear rotators, we therefore remove these objects whose period or cause for varibility is uncertain from our sample, with the exception of two systems for which only the pulsation of the subdwarf is detected, but no rotation of the companion even after pre-whitening is performed. This effectively reduces our sample to 96 systems, out of which 35 show no observed variability due to rotation, and 61 show variability consistent with rotation spun-up by accretion. Considering this sample, the fraction of composite systems with evidence for accretion is $63.5^{+4.6}_{-5.1}$\%, where the uncertainties were calculated assuming a binomial distribution given the low-number statistics \citep[see e.g.][]{burgasser2003} and indicate the 68\% confidence level interval.

However, it is important to notice that the probability of detection of variability is not homogeneous throughout the sample. The NOV systems are, on average, fainter than the systems for which rotational variability has been detected (median $G = 13.4$ for rotational variables, compared to $G = 14.6$ for NOV systems). They are also in fields showing larger crowding, with the median of {\tt CROWDSAP} being equal to 0.97 for rotational variables, but 0.89 for NOV systems. Furthermore, most of the NOV systems have data in only one sector (29 out of 33, 88\%), while almost 30\% of the rotational variables have data in more than one sector. This implies that the detection power is smaller for the systems in which no variability has been detected, compared to those with observed variability.

\begin{figure}
\centering
\includegraphics[width=\hsize]{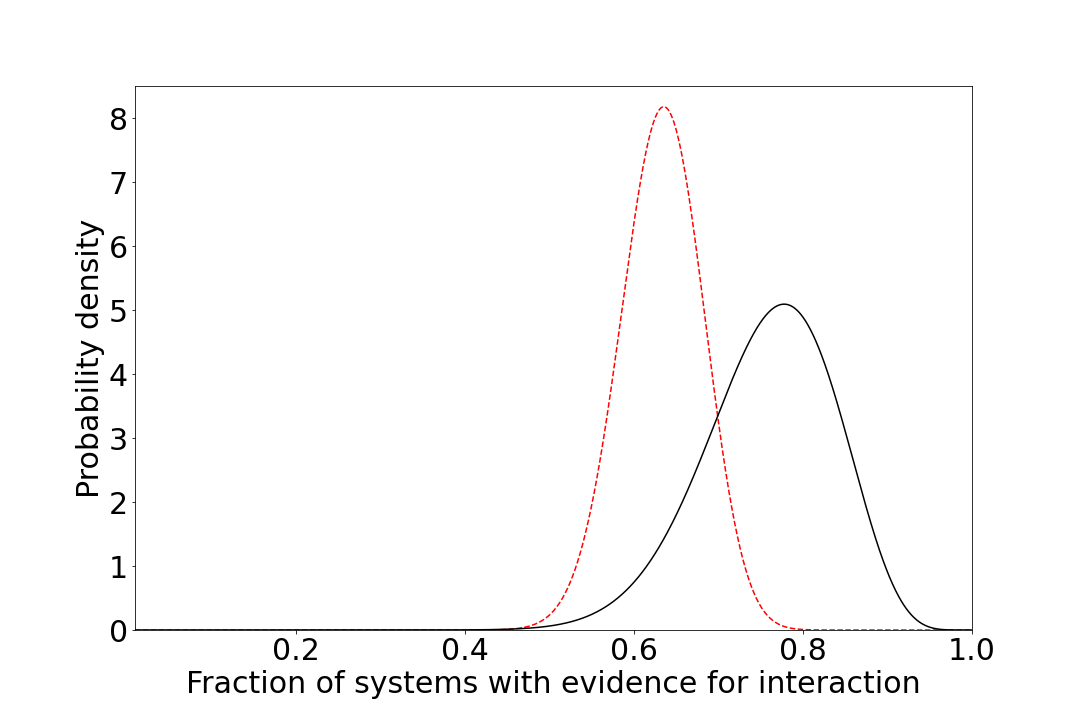}
\caption{Probability density for the fraction of composite systems showing evidence for interaction. The black solid line assumes the reduced sample, excluding stars that are faint and/or in crowded regions. The red dashed line assumes the whole sample.}
\label{binomial}
\end{figure}

To minimise these factors, we estimate the fraction of composite stars with evidence for interaction considering only objects brighter than $G = 13.5$ and with ${\tt CROWDSAP} \geq 0.97$. In this case there are 26 systems, out of 21 are consistent with rotational variables. This yields a fraction of $77.8^{+5.9}_{-9.8}$\%. For a confidence interval of 99.7\% (3$\sigma$), the fraction is as high as 93\% (see Fig.~\ref{binomial}). For the six systems brighter than $G = 12.0$, the fraction of rotational variables is 100\%.

Our search for common proper motion companions has resulted in 16 candidates, with projected orbital distances ranging from 2350 to 18\,300~AU. Considering that 2\,938 stars in our sample have passed the good astrometry criteria, this suggests that only an extremely small fraction of hot subdwarfs are in very wide binaries. Among the progenitors, the fraction of wide binaries is around 15\% \citep{zapatero2004, raghavan2010, moe2017, moe2019}. As the fraction is much smaller for hot subdwarfs, this suggests that they cannot easily be formed in these non-interacting systems, and therefore they would also not be expected to form from single stars.

We note also that for six objects there is evidence of an inner binary companion (see Appendix~\ref{cpm_cand}), suggesting that these systems are rather hierarchical triples. For the progenitors, \citet{elbadry2018} have found that, for 36\% of common proper motion pairs in the main sequence, at least one of the components is a close binary. Combining this with a wide binary fraction of 15\% would suggest a rate of 5.4\% hierarchical triples for progenitor systems. According to \citet{moe2017}, the fraction of triple and quadruple systems for solar-type stars is $10\pm2$\%. The real fraction of triple or multiple systems is unknown for our systems, because no comprehensive search for close companions has been performed for most of them, but the high fraction among the progenitors suggests that the few detected common proper motion pairs could be the progeny of triples, where the inner binary gave origin to the hot subdwarf through one of the canonical binary interaction channels. In fact, \citet{igoshev2020}, who have also found a smaller fraction of wide companions to hot subdwarfs than for their progenitors, suggest that the low fraction of wide companions could be explained by the third member of some systems becoming unbound when the inner binary, which gives origin to the hot subdwarf, goes through a common envelope phase.

\section{Conclusion}
\label{conclusion}

We find that (i) the fraction of hot subdwarfs in composite binaries in which there is evidence for past interaction is higher than 60\%, and can approach 100\% when detection limits are taken into account, and (ii) only an extremely small number of hot subdwarfs (16 out of 2\,938) could be part of a common proper motion pair. We interpret this as a strong evidence against the formation of hot subdwarf stars in single systems, and conclude that binary interaction is in fact always required.

Binary evolution scenarios must, therefore, also explain currently single objects. The merger of two He white dwarfs is one possible channel, although the agreement between observation and model predictions is still poor. The scenario proposed by \citet{demarchi1996}, in which envelope stripping is caused by dynamical processes, would disrupt wide binaries and thus, in principle, also produce single hot subdwarfs with no common proper motion companions. However, the scenario has been proposed for dense globular clusters and would not work for field stars. A promising alternative is offered by the scenario proposed by \citet{clausen2011}, in which single hot subdwarfs form from the merger of a He white dwarf with a hydrogen-burning star. This scenario seems to be able to better explain the observed rotation rates and mass distribution. 

It must be noted, however, that the observed mass distribution of hot subdwarfs is still a work-in-progress, as they have historically been assumed to be have masses equal to the canonical He-core flash value of around $0.5~M\sun$ \citep{heber1986, saffer1994}. Moreover, the apparent lack of companions could be explained by other factors, such as low-mass companions, which can be difficult to detect \citep{nelemans2010}. The possible shortage of progenitors is also not established, since surveys for He-core white dwarfs are still ongoing and the population is growing \citep{pelisoli2019, brown2020, kosakowski2020}. Moreover, alternative non-single formation scenarios must be considered as well, such as supernova stripping \citep{justham2009}, the merger of cataclysmic variables and AM~CVn systems \citep{nelemans2005, zorotovic2017}, or interaction with sub-stellar companions \citep{soker1998}. Upcoming large spectroscopic surveys that will allow precise radial velocity measurements for hot subdwarfs \citep[e.g. 4MOST,][]{4most} will provide a better insight onto the true fraction of currently single hot subdwarfs, as well as onto the observed density of possible progenitors, and further constrain the possible formation scenarios of hot subdwarfs.

\begin{acknowledgements}
We are grateful to Sydney Barnes for an enlightening discussion on stellar rotation that helped motivate this work.

We thank Sydney Barnes and Valerie Van Grootel for providing comments on an earlier version of this manuscript.

IP was partially funded by the Deutsche Forschungsgemeinschaft under grant GE2506/12-1.

This work was supported by a fellowship for postdoctoral researchers from the Alexander von Humboldt Foundation awarded to JV.

ASB gratefully acknowledges financial support from the Polish National Science Center under projects No.\,UMO-2017/26/E/ST9/00703 and UMO-2017/25/B/ST9/02218.

This research made use of Astropy,\footnote{http://www.astropy.org} a community-developed core Python package for Astronomy \citep{astropy:2013, astropy:2018}, and of {\sc TOPCAT} \citep{topcat}.

This paper includes data collected by the TESS mission. Funding for the TESS mission is provided by the NASA Explorer Program.

This work has made use of data from the European Space Agency (ESA) mission
{\it Gaia} (\url{https://www.cosmos.esa.int/gaia}), processed by the {\it Gaia}
Data Processing and Analysis Consortium (DPAC,
\url{https://www.cosmos.esa.int/web/gaia/dpac/consortium}). Funding for the DPAC
has been provided by national institutions, in particular the institutions
participating in the {\it Gaia} Multilateral Agreement.

\end{acknowledgements}


\bibliographystyle{aa} 
\bibliography{TESScomposites} 

\begin{thebibliography}{108}
\expandafter\ifx\csname natexlab\endcsname\relax\def\natexlab#1{#1}\fi

\bibitem[{{Andrews} {et~al.}(2017){Andrews}, {Chanam{\'e}}, \&
  {Ag{\"u}eros}}]{andrews2017}
{Andrews}, J.~J., {Chanam{\'e}}, J., \& {Ag{\"u}eros}, M.~A. 2017, \mnras, 472,
  675

\bibitem[{{Astropy Collaboration} {et~al.}(2013){Astropy Collaboration},
  {Robitaille}, {Tollerud}, {Greenfield}, {Droettboom}, {Bray}, {Aldcroft},
  {Davis}, {Ginsburg}, {Price-Whelan}, {Kerzendorf}, {Conley}, {Crighton},
  {Barbary}, {Muna}, {Ferguson}, {Grollier}, {Parikh}, {Nair}, {Unther},
  {Deil}, {Woillez}, {Conseil}, {Kramer}, {Turner}, {Singer}, {Fox}, {Weaver},
  {Zabalza}, {Edwards}, {Azalee Bostroem}, {Burke}, {Casey}, {Crawford},
  {Dencheva}, {Ely}, {Jenness}, {Labrie}, {Lim}, {Pierfederici}, {Pontzen},
  {Ptak}, {Refsdal}, {Servillat}, \& {Streicher}}]{astropy:2013}
{Astropy Collaboration}, {Robitaille}, T.~P., {Tollerud}, E.~J., {et~al.} 2013,
  \aap, 558, A33

\bibitem[{{Baran} {et~al.}(2009){Baran}, {Oreiro}, {Pigulski}, {P{\'e}rez
  Hern{\'a}ndez}, {Ulla}, {Reed}, {Rodr{\'\i}guez-L{\'o}pez}, {Moskalik},
  {Kim}, {Chen}, {Crowe}, {Siwak}, {Armendarez}, {Binder}, {Choo}, {Dye},
  {Eggen}, {Garrido}, {Gonz{\'a}lez P{\'e}rez}, {Harms}, {Huang}, {Kozie{\l}},
  {Lee}, {MacDonald}, {Fox Machado}, {Monserrat}, {Stevick}, {Stewart},
  {Terry}, {Zhou}, \& {Zo{\l}a}}]{baran2009}
{Baran}, A., {Oreiro}, R., {Pigulski}, A., {et~al.} 2009, \mnras, 392, 1092

\bibitem[{{Baran} {et~al.}(2012){Baran}, {Reed}, {Stello}, {{\O}stensen},
  {Telting}, {Pak{\v{s}}tien{\"e}}, {O'Toole}, {Silvotti}, {Degroote},
  {Bloemen}, {Hu}, {Van Grootel}, {Clarke}, {Van Cleve}, {Thompson}, \&
  {Kawaler}}]{baran2012}
{Baran}, A.~S., {Reed}, M.~D., {Stello}, D., {et~al.} 2012, \mnras, 424, 2686

\bibitem[{{Barlow} {et~al.}(2013){Barlow}, {Liss}, {Wade}, \&
  {Green}}]{barlow2013}
{Barlow}, B.~N., {Liss}, S.~E., {Wade}, R.~A., \& {Green}, E.~M. 2013, \apj,
  771, 23

\bibitem[{{Barlow} {et~al.}(2012){Barlow}, {Wade}, {Liss}, {{\O}stensen}, \&
  {Van Winckel}}]{barlow2012}
{Barlow}, B.~N., {Wade}, R.~A., {Liss}, S.~E., {{\O}stensen}, R.~H., \& {Van
  Winckel}, H. 2012, \apj, 758, 58

\bibitem[{{Basri} \& {Nguyen}(2018)}]{Basri2018}
{Basri}, G. \& {Nguyen}, H.~T. 2018, \apj, 863, 190

\bibitem[{{Brown} {et~al.}(2020){Brown}, {Kilic}, {Kosakowski}, {Andrews},
  {Heinke}, {Ag{\"u}eros}, {Camilo}, {Gianninas}, {Hermes}, \&
  {Kenyon}}]{brown2020}
{Brown}, W.~R., {Kilic}, M., {Kosakowski}, A., {et~al.} 2020, \apj, 889, 49

\bibitem[{{Burgasser} {et~al.}(2003){Burgasser}, {Kirkpatrick}, {Reid},
  {Brown}, {Miskey}, \& {Gizis}}]{burgasser2003}
{Burgasser}, A.~J., {Kirkpatrick}, J.~D., {Reid}, I.~N., {et~al.} 2003, \apj,
  586, 512

\bibitem[{{Charpinet} {et~al.}(2009){Charpinet}, {Brassard}, {Fontaine},
  {Green}, {Van Grootel}, {Randall}, \& {Chayer}}]{charpinet2009}
{Charpinet}, S., {Brassard}, P., {Fontaine}, G., {et~al.} 2009, in American
  Institute of Physics Conference Series, Vol. 1170, American Institute of
  Physics Conference Series, ed. J.~A. {Guzik} \& P.~A. {Bradley}, 585--596

\bibitem[{{Chen} {et~al.}(2013){Chen}, {Han}, {Deca}, \&
  {Podsiadlowski}}]{chen2013}
{Chen}, X., {Han}, Z., {Deca}, J., \& {Podsiadlowski}, P. 2013, \mnras, 434,
  186

\bibitem[{{Clausen} \& {Wade}(2011)}]{clausen2011}
{Clausen}, D. \& {Wade}, R.~A. 2011, \apjl, 733, L42

\bibitem[{{Copperwheat} {et~al.}(2011){Copperwheat}, {Morales-Rueda}, {Marsh},
  {Maxted}, \& {Heber}}]{copperwheat2011}
{Copperwheat}, C.~M., {Morales-Rueda}, L., {Marsh}, T.~R., {Maxted}, P.~F.~L.,
  \& {Heber}, U. 2011, \mnras, 415, 1381

\bibitem[{{Cutri} \& {et al.}(2012)}]{cutri2012}
{Cutri}, R.~M. \& {et al.} 2012, VizieR Online Data Catalog, 2311

\bibitem[{{Davenport} {et~al.}(2019){Davenport}, {Covey}, {Clarke}, {Boeck},
  {Cornet}, \& {Hawley}}]{davenport2019}
{Davenport}, J. R.~A., {Covey}, K.~R., {Clarke}, R.~W., {et~al.} 2019, \apj,
  871, 241

\bibitem[{{D'Cruz} {et~al.}(1996){D'Cruz}, {Dorman}, {Rood}, \&
  {O'Connell}}]{dcruz1996}
{D'Cruz}, N.~L., {Dorman}, B., {Rood}, R.~T., \& {O'Connell}, R.~W. 1996, \apj,
  466, 359

\bibitem[{{de Jong} {et~al.}(2014){de Jong}, {Barden}, {Bellido-Tirado},
  {Brynnel}, {Chiappini}, {Depagne}, {Haynes}, {Johl}, {Phillips}, {Schnurr},
  {Schwope}, {Walcher}, {Bauer}, {Cescutti}, {Cioni}, {Dionies}, {Enke},
  {Haynes}, {Kelz}, {Kitaura}, {Lamer}, {Minchev}, {M{\"u}ller}, {Nuza},
  {Olaya}, {Piffl}, {Popow}, {Saviauk}, {Steinmetz}, {Ural}, {Valentini},
  {Winkler}, {Wisotzki}, {Ansorge}, {Banerji}, {Gonzalez Solares}, {Irwin},
  {Kennicutt}, {King}, {McMahon}, {Koposov}, {Parry}, {Sun}, {Walton},
  {Finger}, {Iwert}, {Krumpe}, {Lizon}, {Mainieri}, {Amans}, {Bonifacio},
  {Cohen}, {Fran{\c c}ois}, {Jagourel}, {Mignot}, {Royer}, {Sartoretti},
  {Bender}, {Hess}, {Lang-Bardl}, {Muschielok}, {Schlichter}, {B{\"o}hringer},
  {Boller}, {Bongiorno}, {Brusa}, {Dwelly}, {Merloni}, {Nandra}, {Salvato},
  {Pragt}, {Navarro}, {Gerlofsma}, {Roelfsema}, {Dalton}, {Middleton}, {Tosh},
  {Boeche}, {Caffau}, {Christlieb}, {Grebel}, {Hansen}, {Koch}, {Ludwig},
  {Mandel}, {Quirrenbach}, {Sbordone}, {Seifert}, {Thimm}, {Helmi}, {trager},
  {Bensby}, {Feltzing}, {Ruchti}, {Edvardsson}, {Korn}, {Lind}, {Boland},
  {Colless}, {Frost}, {Gilbert}, {Gillingham}, {Lawrence}, {Legg}, {Saunders},
  {Sheinis}, {Driver}, {Robotham}, {Bacon}, {Caillier}, {Kosmalski}, {Laurent},
  \& {Richard}}]{4most}
{de Jong}, R.~S., {Barden}, S., {Bellido-Tirado}, O., {et~al.} 2014, in
  \procspie, Vol. 9147, Ground-based and Airborne Instrumentation for Astronomy
  V, 91470M

\bibitem[{{de Marchi} \& {Paresce}(1996)}]{demarchi1996}
{de Marchi}, G. \& {Paresce}, F. 1996, \apj, 467, 658

\bibitem[{{De Rosa} {et~al.}(2014){De Rosa}, {Patience}, {Wilson}, {Schneider},
  {Wiktorowicz}, {Vigan}, {Marois}, {Song}, {Macintosh}, {Graham}, {Doyon},
  {Bessell}, {Thomas}, \& {Lai}}]{derosa2014}
{De Rosa}, R.~J., {Patience}, J., {Wilson}, P.~A., {et~al.} 2014, \mnras, 437,
  1216

\bibitem[{{Deca} {et~al.}(2012){Deca}, {Marsh}, {{\O}stensen}, {Morales-Rueda},
  {Copperwheat}, {Wade}, {Stark}, {Maxted}, {Nelemans}, \& {Heber}}]{deca2012}
{Deca}, J., {Marsh}, T.~R., {{\O}stensen}, R.~H., {et~al.} 2012, \mnras, 421,
  2798

\bibitem[{{Deschamps} {et~al.}(2013){Deschamps}, {Siess}, {Davis}, \&
  {Jorissen}}]{deschamps2013}
{Deschamps}, R., {Siess}, L., {Davis}, P.~J., \& {Jorissen}, A. 2013, \aap,
  557, A40

\bibitem[{{Drilling} {et~al.}(2013){Drilling}, {Jeffery}, {Heber}, {Moehler},
  \& {Napiwotzki}}]{drilling2013}
{Drilling}, J.~S., {Jeffery}, C.~S., {Heber}, U., {Moehler}, S., \&
  {Napiwotzki}, R. 2013, \aap, 551, A31

\bibitem[{{Duch{\^e}ne} \& {Kraus}(2013)}]{duchene2013}
{Duch{\^e}ne}, G. \& {Kraus}, A. 2013, \araa, 51, 269

\bibitem[{{El-Badry} \& {Rix}(2018)}]{elbadry2018}
{El-Badry}, K. \& {Rix}, H.-W. 2018, \mnras, 480, 4884

\bibitem[{{El-Badry} \& {Rix}(2019)}]{elbadry2019}
{El-Badry}, K. \& {Rix}, H.-W. 2019, \mnras, 482, L139

\bibitem[{{Evans} {et~al.}(2018){Evans}, {Riello}, {De Angeli}, {Carrasco},
  {Montegriffo}, {Fabricius}, {Jordi}, {Palaversa}, {Diener}, {Busso},
  {Cacciari}, {van Leeuwen}, {Burgess}, {Davidson}, {Harrison}, {Hodgkin},
  {Pancino}, {Richards}, {Altavilla}, {Balaguer-N{\'u}{\~n}ez}, {Barstow},
  {Bellazzini}, {Brown}, {Castellani}, {Cocozza}, {De Luise}, {Delgado},
  {Ducourant}, {Galleti}, {Gilmore}, {Giuffrida}, {Holl}, {Kewley}, {Koposov},
  {Marinoni}, {Marrese}, {Osborne}, {Piersimoni}, {Portell}, {Pulone},
  {Ragaini}, {Sanna}, {Terrett}, {Walton}, {Wevers}, \&
  {Wyrzykowski}}]{evans2018}
{Evans}, D.~W., {Riello}, M., {De Angeli}, F., {et~al.} 2018, \aap, 616, A4

\bibitem[{{Fouesneau} {et~al.}(2019){Fouesneau}, {Rix}, {von Hippel}, {Hogg},
  \& {Tian}}]{fouesneau2019}
{Fouesneau}, M., {Rix}, H.-W., {von Hippel}, T., {Hogg}, D.~W., \& {Tian}, H.
  2019, \apj, 870, 9

\bibitem[{{Gaia Collaboration} {et~al.}(2018){Gaia Collaboration}, {Brown},
  {Vallenari}, {Prusti}, {de Bruijne}, {Babusiaux}, {Bailer-Jones}, {Biermann},
  {Evans}, {Eyer}, {Jansen}, {Jordi}, {Klioner}, {Lammers}, {Lindegren},
  {Luri}, {Mignard}, {Panem}, {Pourbaix}, {Randich}, {Sartoretti}, {Siddiqui},
  {Soubiran}, {van Leeuwen}, {Walton}, {Arenou}, {Bastian}, {Cropper},
  {Drimmel}, {Katz}, {Lattanzi}, {Bakker}, {Cacciari}, {Casta{\~n}eda},
  {Chaoul}, {Cheek}, {De Angeli}, {Fabricius}, {Guerra}, {Holl}, {Masana},
  {Messineo}, {Mowlavi}, {Nienartowicz}, {Panuzzo}, {Portell}, {Riello},
  {Seabroke}, {Tanga}, {Th{\'e}venin}, {Gracia-Abril}, {Comoretto},
  {Garcia-Reinaldos}, {Teyssier}, {Altmann}, {Andrae}, {Audard},
  {Bellas-Velidis}, {Benson}, {Berthier}, {Blomme}, {Burgess}, {Busso},
  {Carry}, {Cellino}, {Clementini}, {Clotet}, {Creevey}, {Davidson}, {De
  Ridder}, {Delchambre}, {Dell'Oro}, {Ducourant},
  {Fern{\'a}ndez-Hern{\'a}ndez}, {Fouesneau}, {Fr{\'e}mat}, {Galluccio},
  {Garc{\'\i}a-Torres}, {Gonz{\'a}lez-N{\'u}{\~n}ez}, {Gonz{\'a}lez-Vidal},
  {Gosset}, {Guy}, {Halbwachs}, {Hambly}, {Harrison}, {Hern{\'a}ndez},
  {Hestroffer}, {Hodgkin}, {Hutton}, {Jasniewicz}, {Jean-Antoine-Piccolo},
  {Jordan}, {Korn}, {Krone-Martins}, {Lanzafame}, {Lebzelter}, {L{\"o}ffler},
  {Manteiga}, {Marrese}, {Mart{\'\i}n-Fleitas}, {Moitinho}, {Mora}, {Muinonen},
  {Osinde}, {Pancino}, {Pauwels}, {Petit}, {Recio-Blanco}, {Richards},
  {Rimoldini}, {Robin}, {Sarro}, {Siopis}, {Smith}, {Sozzetti}, {S{\"u}veges},
  {Torra}, {van Reeven}, {Abbas}, {Abreu Aramburu}, {Accart}, {Aerts},
  {Altavilla}, {{\'A}lvarez}, {Alvarez}, {Alves}, {Anderson}, {Andrei},
  {Anglada Varela}, {Antiche}, {Antoja}, {Arcay}, {Astraatmadja}, {Bach},
  {Baker}, {Balaguer-N{\'u}{\~n}ez}, {Balm}, {Barache}, {Barata}, {Barbato},
  {Barblan}, {Barklem}, {Barrado}, {Barros}, {Barstow}, {Bartholom{\'e}
  Mu{\~n}oz}, {Bassilana}, {Becciani}, {Bellazzini}, {Berihuete}, {Bertone},
  {Bianchi}, {Bienaym{\'e}}, {Blanco-Cuaresma}, {Boch}, {Boeche}, {Bombrun},
  {Borrachero}, {Bossini}, {Bouquillon}, {Bourda}, {Bragaglia}, {Bramante},
  {Breddels}, {Bressan}, {Brouillet}, {Br{\"u}semeister}, {Brugaletta},
  {Bucciarelli}, {Burlacu}, {Busonero}, {Butkevich}, {Buzzi}, {Caffau},
  {Cancelliere}, {Cannizzaro}, {Cantat-Gaudin}, {Carballo}, {Carlucci},
  {Carrasco}, {Casamiquela}, {Castellani}, {Castro-Ginard}, {Charlot},
  {Chemin}, {Chiavassa}, {Cocozza}, {Costigan}, {Cowell}, {Crifo}, {Crosta},
  {Crowley}, {Cuypers}, {Dafonte}, {Damerdji}, {Dapergolas}, {David}, {David},
  {de Laverny}, {De Luise}, {De March}, {de Martino}, {de Souza}, {de Torres},
  {Debosscher}, {del Pozo}, {Delbo}, {Delgado}, {Delgado}, {Di Matteo},
  {Diakite}, {Diener}, {Distefano}, {Dolding}, {Drazinos}, {Dur{\'a}n},
  {Edvardsson}, {Enke}, {Eriksson}, {Esquej}, {Eynard Bontemps}, {Fabre},
  {Fabrizio}, {Faigler}, {Falc{\~a}o}, {Farr{\`a}s Casas}, {Federici},
  {Fedorets}, {Fernique}, {Figueras}, {Filippi}, {Findeisen}, {Fonti},
  {Fraile}, {Fraser}, {Fr{\'e}zouls}, {Gai}, {Galleti}, {Garabato},
  {Garc{\'\i}a-Sedano}, {Garofalo}, {Garralda}, {Gavel}, {Gavras}, {Gerssen},
  {Geyer}, {Giacobbe}, {Gilmore}, {Girona}, {Giuffrida}, {Glass}, {Gomes},
  {Granvik}, {Gueguen}, {Guerrier}, {Guiraud}, {Guti{\'e}rrez-S{\'a}nchez},
  {Haigron}, {Hatzidimitriou}, {Hauser}, {Haywood}, {Heiter}, {Helmi}, {Heu},
  {Hilger}, {Hobbs}, {Hofmann}, {Holland}, {Huckle}, {Hypki}, {Icardi},
  {Jan{\ss}en}, {Jevardat de Fombelle}, {Jonker}, {Juh{\'a}sz}, {Julbe},
  {Karampelas}, {Kewley}, {Klar}, {Kochoska}, {Kohley}, {Kolenberg},
  {Kontizas}, {Kontizas}, {Koposov}, {Kordopatis}, {Kostrzewa-Rutkowska},
  {Koubsky}, {Lambert}, {Lanza}, {Lasne}, {Lavigne}, {Le Fustec}, {Le
  Poncin-Lafitte}, {Lebreton}, {Leccia}, {Leclerc}, {Lecoeur-Taibi},
  {Lenhardt}, {Leroux}, {Liao}, {Licata}, {Lindstr{\o}m}, {Lister}, {Livanou},
  {Lobel}, {L{\'o}pez}, {Managau}, {Mann}, {Mantelet}, {Marchal}, {Marchant},
  {Marconi}, {Marinoni}, {Marschalk{\'o}}, {Marshall}, {Martino}, {Marton},
  {Mary}, {Massari}, {Matijevi{\v{c}}}, {Mazeh}, {McMillan}, {Messina},
  {Michalik}, {Millar}, {Molina}, {Molinaro}, {Moln{\'a}r}, {Montegriffo},
  {Mor}, {Morbidelli}, {Morel}, {Morris}, {Mulone}, {Muraveva}, {Musella},
  {Nelemans}, {Nicastro}, {Noval}, {O'Mullane}, {Ord{\'e}novic},
  {Ord{\'o}{\~n}ez-Blanco}, {Osborne}, {Pagani}, {Pagano}, {Pailler},
  {Palacin}, {Palaversa}, {Panahi}, {Pawlak}, {Piersimoni}, {Pineau}, {Plachy},
  {Plum}, {Poggio}, {Poujoulet}, {Pr{\v{s}}a}, {Pulone}, {Racero}, {Ragaini},
  {Rambaux}, {Ramos-Lerate}, {Regibo}, {Reyl{\'e}}, {Riclet}, {Ripepi}, {Riva},
  {Rivard}, {Rixon}, {Roegiers}, {Roelens}, {Romero-G{\'o}mez}, {Rowell},
  {Royer}, {Ruiz-Dern}, {Sadowski}, {Sagrist{\`a} Sell{\'e}s}, {Sahlmann},
  {Salgado}, {Salguero}, {Sanna}, {Santana-Ros}, {Sarasso}, {Savietto},
  {Schultheis}, {Sciacca}, {Segol}, {Segovia}, {S{\'e}gransan}, {Shih},
  {Siltala}, {Silva}, {Smart}, {Smith}, {Solano}, {Solitro}, {Sordo}, {Soria
  Nieto}, {Souchay}, {Spagna}, {Spoto}, {Stampa}, {Steele},
  {Steidelm{\"u}ller}, {Stephenson}, {Stoev}, {Suess}, {Surdej}, {Szabados},
  {Szegedi-Elek}, {Tapiador}, {Taris}, {Tauran}, {Taylor}, {Teixeira},
  {Terrett}, {Teyssand ier}, {Thuillot}, {Titarenko}, {Torra Clotet}, {Turon},
  {Ulla}, {Utrilla}, {Uzzi}, {Vaillant}, {Valentini}, {Valette}, {van Elteren},
  {Van Hemelryck}, {van Leeuwen}, {Vaschetto}, {Vecchiato}, {Veljanoski},
  {Viala}, {Vicente}, {Vogt}, {von Essen}, {Voss}, {Votruba}, {Voutsinas},
  {Walmsley}, {Weiler}, {Wertz}, {Wevers}, {Wyrzykowski}, {Yoldas},
  {{\v{Z}}erjal}, {Ziaeepour}, {Zorec}, {Zschocke}, {Zucker}, {Zurbach}, \&
  {Zwitter}}]{gaiadr2}
{Gaia Collaboration}, {Brown}, A.~G.~A., {Vallenari}, A., {et~al.} 2018, \aap,
  616, A1

\bibitem[{{Geier}(2020)}]{geier2020}
{Geier}, S. 2020, arXiv e-prints, arXiv:2002.10896

\bibitem[{{Geier} {et~al.}(2011{\natexlab{a}}){Geier}, {Classen}, \&
  {Heber}}]{geier2011b}
{Geier}, S., {Classen}, L., \& {Heber}, U. 2011{\natexlab{a}}, \apjl, 733, L13

\bibitem[{{Geier} {et~al.}(2015){Geier}, {F{\"u}rst}, {Ziegerer}, {Kupfer},
  {Heber}, {Irrgang}, {Wang}, {Liu}, {Han}, {Sesar}, {Levitan}, {Kotak},
  {Magnier}, {Smith}, {Burgett}, {Chambers}, {Flewelling}, {Kaiser},
  {Wainscoat}, \& {Waters}}]{geier2015}
{Geier}, S., {F{\"u}rst}, F., {Ziegerer}, E., {et~al.} 2015, Science, 347, 1126

\bibitem[{{Geier} \& {Heber}(2012)}]{geier2012}
{Geier}, S. \& {Heber}, U. 2012, \aap, 543, A149

\bibitem[{{Geier} {et~al.}(2013){Geier}, {Heber}, {Heuser}, {Classen},
  {O'Toole}, \& {Edelmann}}]{geier2013}
{Geier}, S., {Heber}, U., {Heuser}, C., {et~al.} 2013, \aap, 551, L4

\bibitem[{{Geier} {et~al.}(2011{\natexlab{b}}){Geier}, {Hirsch}, {Tillich},
  {Maxted}, {Bentley}, {{\O}stensen}, {Heber}, {G{\"a}nsicke}, {Marsh},
  {Napiwotzki}, {Barlow}, \& {O'Toole}}]{geier2011}
{Geier}, S., {Hirsch}, H., {Tillich}, A., {et~al.} 2011{\natexlab{b}}, \aap,
  530, A28

\bibitem[{{Geier} {et~al.}(2017){Geier}, {{\O}stensen}, {Nemeth}, {Gentile
  Fusillo}, {G{\"a}nsicke}, {Telting}, {Green}, \& {Schaffenroth}}]{geier2017}
{Geier}, S., {{\O}stensen}, R.~H., {Nemeth}, P., {et~al.} 2017, \aap, 600, A50

\bibitem[{{Geier} {et~al.}(2019){Geier}, {Raddi}, {Gentile Fusillo}, \&
  {Marsh}}]{geier2019}
{Geier}, S., {Raddi}, R., {Gentile Fusillo}, N.~P., \& {Marsh}, T.~R. 2019,
  \aap, 621, A38

\bibitem[{{Green} {et~al.}(2003){Green}, {Fontaine}, {Reed}, {Callerame},
  {Seitenzahl}, {White}, {Hyde}, {{\O}stensen}, {Cordes}, {Brassard}, {Falter},
  {Jeffery}, {Dreizler}, {Schuh}, {Giovanni}, {Edelmann}, {Rigby}, \&
  {Bronowska}}]{green2003}
{Green}, E.~M., {Fontaine}, G., {Reed}, M.~D., {et~al.} 2003, \apjl, 583, L31

\bibitem[{{Han} {et~al.}(2003){Han}, {Podsiadlowski}, {Maxted}, \&
  {Marsh}}]{han2003}
{Han}, Z., {Podsiadlowski}, P., {Maxted}, P.~F.~L., \& {Marsh}, T.~R. 2003,
  \mnras, 341, 669

\bibitem[{{Han} {et~al.}(2002){Han}, {Podsiadlowski}, {Maxted}, {Marsh}, \&
  {Ivanova}}]{han2002}
{Han}, Z., {Podsiadlowski}, P., {Maxted}, P.~F.~L., {Marsh}, T.~R., \&
  {Ivanova}, N. 2002, \mnras, 336, 449

\bibitem[{{Heber}(1986)}]{heber1986}
{Heber}, U. 1986, \aap, 155, 33

\bibitem[{{Heber}(2016)}]{heber2016}
{Heber}, U. 2016, \pasp, 128, 082001

\bibitem[{{Heber} {et~al.}(2002){Heber}, {Moehler}, {Napiwotzki}, {Thejll}, \&
  {Green}}]{heber2002}
{Heber}, U., {Moehler}, S., {Napiwotzki}, R., {Thejll}, P., \& {Green}, E.~M.
  2002, \aap, 383, 938

\bibitem[{{Henden} {et~al.}(2015){Henden}, {Levine}, {Terrell}, \&
  {Welch}}]{henden2015}
{Henden}, A.~A., {Levine}, S., {Terrell}, D., \& {Welch}, D.~L. 2015, in
  American Astronomical Society Meeting Abstracts, Vol. 225, American
  Astronomical Society Meeting Abstracts \#225, 336.16

\bibitem[{{Howell} {et~al.}(2014){Howell}, {Sobeck}, {Haas}, {Still},
  {Barclay}, {Mullally}, {Troeltzsch}, {Aigrain}, {Bryson}, {Caldwell},
  {Chaplin}, {Cochran}, {Huber}, {Marcy}, {Miglio}, {Najita}, {Smith},
  {Twicken}, \& {Fortney}}]{howell2014}
{Howell}, S.~B., {Sobeck}, C., {Haas}, M., {et~al.} 2014, \pasp, 126, 398

\bibitem[{{Humason} \& {Zwicky}(1947)}]{humason1947}
{Humason}, M.~L. \& {Zwicky}, F. 1947, Contributions from the Mount Wilson
  Observatory / Carnegie Institution of Washington, 724, 1

\bibitem[{{Igoshev} {et~al.}(2020){Igoshev}, {Perets}, \&
  {Michaely}}]{igoshev2020}
{Igoshev}, A.~P., {Perets}, H.~B., \& {Michaely}, E. 2020, \mnras, 494, 1448

\bibitem[{{Justham} {et~al.}(2009){Justham}, {Wolf}, {Podsiadlowski}, \&
  {Han}}]{justham2009}
{Justham}, S., {Wolf}, C., {Podsiadlowski}, P., \& {Han}, Z. 2009, \aap, 493,
  1081

\bibitem[{{Kilkenny} {et~al.}(1997){Kilkenny}, {O'Donoghue}, {Koen}, {Stobie},
  \& {Chen}}]{kilkenny1997}
{Kilkenny}, D., {O'Donoghue}, D., {Koen}, C., {Stobie}, R.~S., \& {Chen}, A.
  1997, \mnras, 287, 867

\bibitem[{{Kippenhahn} \& {Meyer-Hofmeister}(1977)}]{kippenhahn1977}
{Kippenhahn}, R. \& {Meyer-Hofmeister}, E. 1977, \aap, 54, 539

\bibitem[{{Kosakowski} {et~al.}(2020){Kosakowski}, {Kilic}, {Brown}, \&
  {Gianninas}}]{kosakowski2020}
{Kosakowski}, A., {Kilic}, M., {Brown}, W.~R., \& {Gianninas}, A. 2020, arXiv
  e-prints, arXiv:2004.04202

\bibitem[{{Kupfer} {et~al.}(2018){Kupfer}, {Korol}, {Shah}, {Nelemans},
  {Marsh}, {Ramsay}, {Groot}, {Steeghs}, \& {Rossi}}]{kupfer2018}
{Kupfer}, T., {Korol}, V., {Shah}, S., {et~al.} 2018, \mnras, 480, 302

\bibitem[{{Kurucz}(1979)}]{kurucz1979}
{Kurucz}, R.~L. 1979, \apjs, 40, 1

\bibitem[{{Lallement} {et~al.}(2019){Lallement}, {Babusiaux}, {Vergely},
  {Katz}, {Arenou}, {Valette}, {Hottier}, \& {Capitanio}}]{lallement2019}
{Lallement}, R., {Babusiaux}, C., {Vergely}, J.~L., {et~al.} 2019, \aap, 625,
  A135

\bibitem[{{Latour} {et~al.}(2018){Latour}, {Randall}, {Calamida}, {Geier}, \&
  {Moehler}}]{latour2018}
{Latour}, M., {Randall}, S.~K., {Calamida}, A., {Geier}, S., \& {Moehler}, S.
  2018, \aap, 618, A15

\bibitem[{{Leiner} {et~al.}(2018){Leiner}, {Mathieu}, {Gosnell}, \&
  {Sills}}]{leiner2018}
{Leiner}, E., {Mathieu}, R.~D., {Gosnell}, N.~M., \& {Sills}, A. 2018, \apjl,
  869, L29

\bibitem[{{Lindegren} {et~al.}(2018){Lindegren}, {Hern{\'a}ndez}, {Bombrun},
  {Klioner}, {Bastian}, {Ramos-Lerate}, {de Torres}, {Steidelm{\"u}ller},
  {Stephenson}, {Hobbs}, {Lammers}, {Biermann}, {Geyer}, {Hilger}, {Michalik},
  {Stampa}, {McMillan}, {Casta{\~n}eda}, {Clotet}, {Comoretto}, {Davidson},
  {Fabricius}, {Gracia}, {Hambly}, {Hutton}, {Mora}, {Portell}, {van Leeuwen},
  {Abbas}, {Abreu}, {Altmann}, {Andrei}, {Anglada}, {Balaguer-N{\'u}{\~n}ez},
  {Barache}, {Becciani}, {Bertone}, {Bianchi}, {Bouquillon}, {Bourda},
  {Br{\"u}semeister}, {Bucciarelli}, {Busonero}, {Buzzi}, {Cancelliere},
  {Carlucci}, {Charlot}, {Cheek}, {Crosta}, {Crowley}, {de Bruijne}, {de
  Felice}, {Drimmel}, {Esquej}, {Fienga}, {Fraile}, {Gai}, {Garralda},
  {Gonz{\'a}lez-Vidal}, {Guerra}, {Hauser}, {Hofmann}, {Holl}, {Jordan},
  {Lattanzi}, {Lenhardt}, {Liao}, {Licata}, {Lister}, {L{\"o}ffler},
  {Marchant}, {Martin-Fleitas}, {Messineo}, {Mignard}, {Morbidelli}, {Poggio},
  {Riva}, {Rowell}, {Salguero}, {Sarasso}, {Sciacca}, {Siddiqui}, {Smart},
  {Spagna}, {Steele}, {Taris}, {Torra}, {van Elteren}, {van Reeven}, \&
  {Vecchiato}}]{lindegren2018}
{Lindegren}, L., {Hern{\'a}ndez}, J., {Bombrun}, A., {et~al.} 2018, \aap, 616,
  A2

\bibitem[{{Lomb}(1976)}]{lomb1976}
{Lomb}, N.~R. 1976, \apss, 39, 447

\bibitem[{{Maxted} {et~al.}(2001){Maxted}, {Heber}, {Marsh}, \&
  {North}}]{maxted2001}
{Maxted}, P.~F.~L., {Heber}, U., {Marsh}, T.~R., \& {North}, R.~C. 2001,
  \mnras, 326, 1391

\bibitem[{{Maxted} {et~al.}(2000){Maxted}, {Marsh}, \& {North}}]{maxted2000}
{Maxted}, P.~F.~L., {Marsh}, T.~R., \& {North}, R.~C. 2000, \mnras, 317, L41

\bibitem[{{McNamara} {et~al.}(2012){McNamara}, {Jackiewicz}, \&
  {McKeever}}]{mcnamara2012}
{McNamara}, B.~J., {Jackiewicz}, J., \& {McKeever}, J. 2012, \aj, 143, 101

\bibitem[{{Moe} \& {Di Stefano}(2017)}]{moe2017}
{Moe}, M. \& {Di Stefano}, R. 2017, \apjs, 230, 15

\bibitem[{{Moe} {et~al.}(2019){Moe}, {Kratter}, \& {Badenes}}]{moe2019}
{Moe}, M., {Kratter}, K.~M., \& {Badenes}, C. 2019, \apj, 875, 61

\bibitem[{{Morales-Rueda} {et~al.}(2003){Morales-Rueda}, {Maxted}, {Marsh},
  {North}, \& {Heber}}]{morales2003}
{Morales-Rueda}, L., {Maxted}, P.~F.~L., {Marsh}, T.~R., {North}, R.~C., \&
  {Heber}, U. 2003, \mnras, 338, 752

\bibitem[{{Nelemans}(2005)}]{nelemans2005}
{Nelemans}, G. 2005, Astronomical Society of the Pacific Conference Series,
  Vol. 330, {AM CVn stars}, ed. J.~M. {Hameury} \& J.~P. {Lasota}, 27

\bibitem[{{Nelemans}(2010)}]{nelemans2010}
{Nelemans}, G. 2010, \apss, 329, 25

\bibitem[{{N{\'e}meth} {et~al.}(2012){N{\'e}meth}, {Kawka}, \&
  {Vennes}}]{nemeth2012}
{N{\'e}meth}, P., {Kawka}, A., \& {Vennes}, S. 2012, \mnras, 427, 2180

\bibitem[{{O'Connell}(1999)}]{oconnell1999}
{O'Connell}, R.~W. 1999, \araa, 37, 603

\bibitem[{{{\O}stensen} {et~al.}(2010){{\O}stensen}, {Silvotti}, {Charpinet},
  {Oreiro}, {Handler}, {Green}, {Bloemen}, {Heber}, {G{\"a}nsicke}, {Marsh},
  {Kurtz}, {Telting}, {Reed}, {Kawaler}, {Aerts}, {Rodr{\'\i}guez-L{\'o}pez},
  {Vu{\v{c}}kovi{\'c}}, {Ottosen}, {Liimets}, {Quint}, {Van Grootel},
  {Randall}, {Gilliland }, {Kjeldsen}, {Christensen-Dalsgaard}, {Borucki},
  {Koch}, \& {Quintana}}]{ostensen2010}
{{\O}stensen}, R.~H., {Silvotti}, R., {Charpinet}, S., {et~al.} 2010, \mnras,
  409, 1470

\bibitem[{{Pablo} {et~al.}(2012){Pablo}, {Kawaler}, {Reed}, {Bloemen},
  {Charpinet}, {Hu}, {Telting}, {{\O}stensen}, {Baran}, {Green}, {Hermes},
  {Barclay}, {O'Toole}, {Mullally}, {Kurtz}, {Christensen-Dalsgaard},
  {Caldwell}, {Christiansen}, \& {Kinemuchi}}]{pablo2012}
{Pablo}, H., {Kawaler}, S.~D., {Reed}, M.~D., {et~al.} 2012, \mnras, 422, 1343

\bibitem[{{Paczynski}(1991)}]{paczynski1991}
{Paczynski}, B. 1991, \apj, 370, 597

\bibitem[{{Pelisoli} \& {Vos}(2019)}]{pelisoli2019}
{Pelisoli}, I. \& {Vos}, J. 2019, \mnras, 488, 2892

\bibitem[{{Popham} \& {Narayan}(1991)}]{popham1991}
{Popham}, R. \& {Narayan}, R. 1991, \apj, 370, 604

\bibitem[{{Price-Whelan} {et~al.}(2018){Price-Whelan}, {Sip{\H{o}}cz},
  {G{\"u}nther}, {Lim}, {Crawford}, {Conseil}, {Shupe}, {Craig}, {Dencheva},
  {Ginsburg}, {VanderPlas}, {Bradley}, {P{\'e}rez-Su{\'a}rez}, {de Val-Borro},
  {Paper Contributors}, {Aldcroft}, {Cruz}, {Robitaille}, {Tollerud},
  {Coordination Committee}, {Ardelean}, {Babej}, {Bach}, {Bachetti}, {Bakanov},
  {Bamford}, {Barentsen}, {Barmby}, {Baumbach}, {Berry}, {Biscani}, {Boquien},
  {Bostroem}, {Bouma}, {Brammer}, {Bray}, {Breytenbach}, {Buddelmeijer},
  {Burke}, {Calderone}, {Cano Rodr{\'\i}guez}, {Cara}, {Cardoso}, {Cheedella},
  {Copin}, {Corrales}, {Crichton}, {D{\textquoteright}Avella}, {Deil},
  {Depagne}, {Dietrich}, {Donath}, {Droettboom}, {Earl}, {Erben}, {Fabbro},
  {Ferreira}, {Finethy}, {Fox}, {Garrison}, {Gibbons}, {Goldstein}, {Gommers},
  {Greco}, {Greenfield}, {Groener}, {Grollier}, {Hagen}, {Hirst}, {Homeier},
  {Horton}, {Hosseinzadeh}, {Hu}, {Hunkeler}, {Ivezi{\'c}}, {Jain}, {Jenness},
  {Kanarek}, {Kendrew}, {Kern}, {Kerzendorf}, {Khvalko}, {King}, {Kirkby},
  {Kulkarni}, {Kumar}, {Lee}, {Lenz}, {Littlefair}, {Ma}, {Macleod},
  {Mastropietro}, {McCully}, {Montagnac}, {Morris}, {Mueller}, {Mumford},
  {Muna}, {Murphy}, {Nelson}, {Nguyen}, {Ninan}, {N{\"o}the}, {Ogaz}, {Oh},
  {Parejko}, {Parley}, {Pascual}, {Patil}, {Patil}, {Plunkett}, {Prochaska},
  {Rastogi}, {Reddy Janga}, {Sabater}, {Sakurikar}, {Seifert}, {Sherbert},
  {Sherwood-Taylor}, {Shih}, {Sick}, {Silbiger}, {Singanamalla}, {Singer},
  {Sladen}, {Sooley}, {Sornarajah}, {Streicher}, {Teuben}, {Thomas},
  {Tremblay}, {Turner}, {Terr{\'o}n}, {van Kerkwijk}, {de la Vega}, {Watkins},
  {Weaver}, {Whitmore}, {Woillez}, {Zabalza}, \& {Contributors}}]{astropy:2018}
{Price-Whelan}, A.~M., {Sip{\H{o}}cz}, B.~M., {G{\"u}nther}, H.~M., {et~al.}
  2018, \aj, 156, 123

\bibitem[{{Raghavan} {et~al.}(2010){Raghavan}, {McAlister}, {Henry}, {Latham},
  {Marcy}, {Mason}, {Gies}, {White}, \& {ten Brummelaar}}]{raghavan2010}
{Raghavan}, D., {McAlister}, H.~A., {Henry}, T.~J., {et~al.} 2010, \apjs, 190,
  1

\bibitem[{{Rebull} {et~al.}(2016){Rebull}, {Stauffer}, {Bouvier}, {Cody},
  {Hillenbrand}, {Soderblom}, {Valenti}, {Barrado}, {Bouy}, {Ciardi},
  {Pinsonneault}, {Stassun}, {Micela}, {Aigrain}, {Vrba}, {Somers},
  {Christiansen}, {Gillen}, \& {Collier Cameron}}]{rebull2016}
{Rebull}, L.~M., {Stauffer}, J.~R., {Bouvier}, J., {et~al.} 2016, \aj, 152, 113

\bibitem[{{Rebull} {et~al.}(2018){Rebull}, {Stauffer}, {Cody}, {Hillenbrand },
  {David}, \& {Pinsonneault}}]{rebull2018}
{Rebull}, L.~M., {Stauffer}, J.~R., {Cody}, A.~M., {et~al.} 2018, \aj, 155, 196

\bibitem[{{Reed} {et~al.}(2014){Reed}, {Foster}, {Telting}, {{\O}stensen},
  {Farris}, {Oreiro}, \& {Baran}}]{reed2014}
{Reed}, M.~D., {Foster}, H., {Telting}, J.~H., {et~al.} 2014, \mnras, 440, 3809

\bibitem[{{Reinhold} \& {Gizon}(2015)}]{reinhold2015}
{Reinhold}, T. \& {Gizon}, L. 2015, \aap, 583, A65

\bibitem[{{Reinhold} \& {Hekker}(2020)}]{reinhold2020}
{Reinhold}, T. \& {Hekker}, S. 2020, \aap, 635, A43

\bibitem[{{Reinhold} \& {Reiners}(2013)}]{reinhold2013}
{Reinhold}, T. \& {Reiners}, A. 2013, \aap, 557, A11

\bibitem[{{Retterer} \& {King}(1982)}]{retterer1982}
{Retterer}, J.~M. \& {King}, I.~R. 1982, \apj, 254, 214

\bibitem[{{Ricker} {et~al.}(2015){Ricker}, {Winn}, {Vanderspek}, {Latham},
  {Bakos}, {Bean}, {Berta-Thompson}, {Brown}, {Buchhave}, {Butler}, {Butler},
  {Chaplin}, {Charbonneau}, {Christensen-Dalsgaard}, {Clampin}, {Deming},
  {Doty}, {De Lee}, {Dressing}, {Dunham}, {Endl}, {Fressin}, {Ge}, {Henning},
  {Holman}, {Howard}, {Ida}, {Jenkins}, {Jernigan}, {Johnson}, {Kaltenegger},
  {Kawai}, {Kjeldsen}, {Laughlin}, {Levine}, {Lin}, {Lissauer}, {MacQueen},
  {Marcy}, {McCullough}, {Morton}, {Narita}, {Paegert}, {Palle}, {Pepe},
  {Pepper}, {Quirrenbach}, {Rinehart}, {Sasselov}, {Sato}, {Seager},
  {Sozzetti}, {Stassun}, {Sullivan}, {Szentgyorgyi}, {Torres}, {Udry}, \&
  {Villasenor}}]{ricker2015}
{Ricker}, G.~R., {Winn}, J.~N., {Vanderspek}, R., {et~al.} 2015, Journal of
  Astronomical Telescopes, Instruments, and Systems, 1, 014003

\bibitem[{{Riello} {et~al.}(2018){Riello}, {De Angeli}, {Evans}, {Busso},
  {Hambly}, {Davidson}, {Burgess}, {Montegriffo}, {Osborne}, {Kewley},
  {Carrasco}, {Fabricius}, {Jordi}, {Cacciari}, {van Leeuwen}, \&
  {Holland}}]{riello2018}
{Riello}, M., {De Angeli}, F., {Evans}, D.~W., {et~al.} 2018, \aap, 616, A3

\bibitem[{{Saffer} {et~al.}(1994){Saffer}, {Bergeron}, {Koester}, \&
  {Liebert}}]{saffer1994}
{Saffer}, R.~A., {Bergeron}, P., {Koester}, D., \& {Liebert}, J. 1994, \apj,
  432, 351

\bibitem[{{Saffer} {et~al.}(1998){Saffer}, {Livio}, \&
  {Yungelson}}]{saffer1998}
{Saffer}, R.~A., {Livio}, M., \& {Yungelson}, L.~R. 1998, \apj, 502, 394

\bibitem[{{Scargle}(1982)}]{scargle1982}
{Scargle}, J.~D. 1982, \apj, 263, 835

\bibitem[{Schneider {et~al.}(2019)Schneider, Heber, Geier, Latour, \&
  Irrgang}]{schneider_david_2019_3428841}
Schneider, D., Heber, U., Geier, S., Latour, M., \& Irrgang, A. 2019,
  {Fundamental Parameters of Hot Subdwarf Stars from Gaia Astrometry}

\bibitem[{{Silvotti} {et~al.}(2000){Silvotti}, {Solheim}, {Gonzalez Perez},
  {Heber}, {Dreizler}, {Edelmann}, {{\O}stensen}, \& {Kotak}}]{silvotti2000}
{Silvotti}, R., {Solheim}, J.~E., {Gonzalez Perez}, J.~M., {et~al.} 2000, \aap,
  359, 1068

\bibitem[{{Skrutskie} {et~al.}(2006){Skrutskie}, {Cutri}, {Stiening},
  {Weinberg}, {Schneider}, {Carpenter}, {Beichman}, {Capps}, {Chester},
  {Elias}, {Huchra}, {Liebert}, {Lonsdale}, {Monet}, {Price}, {Seitzer},
  {Jarrett}, {Kirkpatrick}, {Gizis}, {Howard}, {Evans}, {Fowler}, {Fullmer},
  {Hurt}, {Light}, {Kopan}, {Marsh}, {McCallon}, {Tam}, {Van Dyk}, \&
  {Wheelock}}]{skrutskie2006}
{Skrutskie}, M.~F., {Cutri}, R.~M., {Stiening}, R., {et~al.} 2006, \aj, 131,
  1163

\bibitem[{{Soker}(1998)}]{soker1998}
{Soker}, N. 1998, \aj, 116, 1308

\bibitem[{{Stark} \& {Wade}(2003)}]{stark2003}
{Stark}, M.~A. \& {Wade}, R.~A. 2003, \aj, 126, 1455

\bibitem[{{Sweigart}(1997)}]{sweigart1997}
{Sweigart}, A.~V. 1997, \apjl, 474, L23

\bibitem[{{Taylor}(2005)}]{topcat}
{Taylor}, M.~B. 2005, in Astronomical Society of the Pacific Conference Series,
  Vol. 347, Astronomical Data Analysis Software and Systems XIV, ed.
  P.~{Shopbell}, M.~{Britton}, \& R.~{Ebert}, 29

\bibitem[{{Tillich} {et~al.}(2011){Tillich}, {Heber}, {Geier}, {Hirsch},
  {Maxted}, {G{\"a}nsicke}, {Marsh}, {Napiwotzki}, {{\O}stensen}, \&
  {Scholz}}]{tillich2011}
{Tillich}, A., {Heber}, U., {Geier}, S., {et~al.} 2011, \aap, 527, A137

\bibitem[{{Toonen} {et~al.}(2012){Toonen}, {Nelemans}, \& {Portegies
  Zwart}}]{toonen2012}
{Toonen}, S., {Nelemans}, G., \& {Portegies Zwart}, S. 2012, \aap, 546, A70

\bibitem[{{Vos} {et~al.}(2020){Vos}, {Bobrick}, \& {Vuckovic}}]{vos2020}
{Vos}, J., {Bobrick}, A., \& {Vuckovic}, M. 2020, arXiv e-prints,
  arXiv:2003.05665

\bibitem[{{Vos} {et~al.}(2018{\natexlab{a}}){Vos}, {N{\'e}meth},
  {Vu{\v{c}}kovi{\'c}}, {{\O}stensen}, \& {Parsons}}]{vos2018}
{Vos}, J., {N{\'e}meth}, P., {Vu{\v{c}}kovi{\'c}}, M., {{\O}stensen}, R., \&
  {Parsons}, S. 2018{\natexlab{a}}, \mnras, 473, 693

\bibitem[{{Vos} {et~al.}(2012){Vos}, {{\O}stensen}, {Degroote}, {De Smedt},
  {Green}, {Heber}, {Van Winckel}, {Acke}, {Bloemen}, {De Cat}, {Exter},
  {Lampens}, {Lombaert}, {Masseron}, {Menu}, {Neyskens}, {Raskin}, {Ringat},
  {Rauch}, {Smolders}, \& {Tkachenko}}]{vos2012}
{Vos}, J., {{\O}stensen}, R.~H., {Degroote}, P., {et~al.} 2012, A\&A, 548, A6

\bibitem[{{Vos} {et~al.}(2013){Vos}, {{\O}stensen}, {N{\'e}meth}, {Green},
  {Heber}, \& {Van Winckel}}]{vos2013}
{Vos}, J., {{\O}stensen}, R.~H., {N{\'e}meth}, P., {et~al.} 2013, A\&A, 559,
  A54

\bibitem[{{Vos} {et~al.}(2017){Vos}, {{\O}stensen}, {Vu{\v c}kovi{\'c}}, \&
  {Van Winckel}}]{vos2017}
{Vos}, J., {{\O}stensen}, R.~H., {Vu{\v c}kovi{\'c}}, M., \& {Van Winckel}, H.
  2017, \aap, 605, A109

\bibitem[{{Vos} {et~al.}(2018{\natexlab{b}}){Vos}, {Zorotovic},
  {Vu{\v{c}}kovi{\'c}}, {Schreiber}, \& {{\O}stensen}}]{vos2018b}
{Vos}, J., {Zorotovic}, M., {Vu{\v{c}}kovi{\'c}}, M., {Schreiber}, M.~R., \&
  {{\O}stensen}, R. 2018{\natexlab{b}}, \mnras, 477, L40

\bibitem[{{Wang} {et~al.}(2018){Wang}, {Gies}, \& {Peters}}]{wang2018}
{Wang}, L., {Gies}, D.~R., \& {Peters}, G.~J. 2018, \apj, 853, 156

\bibitem[{{Weinberg} {et~al.}(1987){Weinberg}, {Shapiro}, \&
  {Wasserman}}]{weinberg1987}
{Weinberg}, M.~D., {Shapiro}, S.~L., \& {Wasserman}, I. 1987, \apj, 312, 367

\bibitem[{{Werner} {et~al.}(2003){Werner}, {Deetjen}, {Dreizler}, {Nagel},
  {Rauch}, \& {Schuh}}]{werner2003}
{Werner}, K., {Deetjen}, J.~L., {Dreizler}, S., {et~al.} 2003, in ASPCS, Vol.
  288, Stellar Atmosphere Modeling, ed. I.~{Hubeny}, D.~{Mihalas}, \&
  K.~{Werner}, 31

\bibitem[{{Zapatero Osorio} \& {Mart{\'\i}n}(2004)}]{zapatero2004}
{Zapatero Osorio}, M.~R. \& {Mart{\'\i}n}, E.~L. 2004, \aap, 419, 167

\bibitem[{{Zhang} {et~al.}(2009){Zhang}, {Chen}, \& {Han}}]{zhang2009}
{Zhang}, X., {Chen}, X., \& {Han}, Z. 2009, \aap, 504, L13

\bibitem[{{Zhao} {et~al.}(2012){Zhao}, {Zhao}, {Chu}, {Jing}, \&
  {Deng}}]{lamost}
{Zhao}, G., {Zhao}, Y., {Chu}, Y., {Jing}, Y., \& {Deng}, L. 2012, arXiv
  e-prints, arXiv:1206.3569

\bibitem[{{Zorotovic} \& {Schreiber}(2017)}]{zorotovic2017}
{Zorotovic}, M. \& {Schreiber}, M.~R. 2017, \mnras, 466, L63

\end{thebibliography}

\onecolumn
\begin{center}
\setlength{\LTcapwidth}{\textwidth} 
\begin{longtable}[c]{r r r r r r r r}
\caption{ The 61 variables in our sample whose variability was attributed to rotation of the main sequence companion, identified by their TESS Input Catalogue (TIC) numbers. $G$ is the magnitude in the {\it Gaia} DR2 catalogue, and the value of {\tt CROWDSAP} is the one calculated by the SPOC pipeline. For objects with data from more than one sector, this is the averaged value. Period and amplitude uncertainties were determined with a hundred Monte-Carlo runs of the multi-component fit to the light curve, re-drawing the fluxes from a normal distribution taking the quoted uncertainties into account. The FAP was calculated using the {\tt astropy} {\tt LombScargle} function. $T\eff$ and radius for the main sequence companion were obtained from a SED fit, as described in the text. Uncertainties are given on the last significant digits, e.g. $3.5805(23) = 3.5805\pm0.0023$, $1.5(2.6) = 1.5\pm2.6$, $6400(400) = 6400\pm400$.} \label{rot} \\
\hline\hline
TIC & $G$ & CROWDSAP & $P$ (days) & Amplitude (\%) & FAP & $T\eff^{\MS}$ (K) & $R^{\MS}$ ($R\sun$)\\
\hline
\endfirsthead
\multicolumn{3}{l}{{\tablename\ \thetable{} continued.}} \\
\hline\hline
TIC & $G$ & CROWDSAP & $P$ (days) & Amplitude (\%) & FAP & $T\eff^{\MS}$ (K) & $R^{MS}$ ($R\sun$)\\
\hline
\endhead
\hline
\endfoot
\hline
\endlastfoot
3990402 & 11.08 &  0.98 & 3.5805(23) & 0.3169(23) & 0.00e+00 & 5794(78) & 1.14(6) \\
12528447 & 12.50 &  1.00 & 0.67138(9) & 0.3340(46) & 0.00e+00 & 6720(163) & 1.61(6) \\
16876025 & 12.86 &  1.00 & 2.326(7) & 0.205(8) & 0.00e+00 & 6172(73) & 1.27(7) \\
30019744 & 12.12 &  0.71 & 2.113(11) & 0.163(7) & 0.00e+00 & 5758(25) & 1.57(6) \\
32556882 & 13.82 &  0.80 & 5.35(5) & 0.454(28) & 0.00e+00 & 5328(20) & 1.221(50) \\
33526769 & 12.24 &  0.99 & 2.2285(41) & 0.1467(36) & 0.00e+00 & 5531(42) & 0.986(40) \\
64112207 & 15.32 &  0.99 & 7.26(16) & 0.252(40) & 5.39e-06 & 5772(75) & 2.10(26) \\
65263746 & 13.40 &  0.78 & 4.498(28) & 0.171(10) & 6.49e-72 & 6146(117) & 0.82(7) \\
68942649 & 10.04 &  0.98 & 2.8383(8) & 0.1954(10) & 0.00e+00 & 6400(400) & 1.10(5) \\
69841801 & 12.25 &  1.00 & 1.251(17) & 0.88(14) & 0.00e+00 & 5896(41) & 1.398(48) \\
70451188 & 14.22 &  0.99 & 2.836(14) & 0.305(19) & 0.00e+00 & 5965(24) & 1.28(6) \\
71248239 & 12.83 &  0.89 & 0.9728(18) & 0.173(9) & 0.00e+00 & 6050(36) & 1.304(34) \\
71716888 & 12.97 &  0.99 & 1.0491(23) & 0.066(29) & 3.75e-24 & 6992(39) & 1.62(8) \\
92865531 & 13.12 &  0.96 & 1.40297(17) & 1.888(7) & 0.00e+00 & 5004(14) & 1.432(37) \\
116416387 & 14.65 &  0.84 & 3.819(38) & 0.402(43) & 0.00e+00 & 5515(46) & 0.99(6) \\
118269334 & 13.82 &  0.97 & 3.073(7) & 0.717(19) & 0.00e+00 & 5479(74) & 0.970(45) \\
143058705 & 14.40 &  0.98 & 5.027(30) & 0.540(28) & 0.00e+00 & 5449(42) & 1.00(6) \\
146323153 & 12.15 &  0.96 & 0.219040(20) & 0.1987(34) & 0.00e+00 & 6793(44) & 1.663(41) \\
149767908 & 13.00 &  0.99 & 0.4557(6) & 0.229(10) & 0.00e+00 & 7300(56) & 1.67(12) \\
151641733 & 12.65 &  0.96 & 0.5592(8) & 0.028(5) & 2.03e-10 & 6540(41) & 1.48(5) \\
157323544 & 12.69 &  0.99 & 4.462(43) & 0.080(7) & 0.00e+00 & 6250(291) & 2.41(10) \\
158335560 & 13.08 &  0.92 & 2.0007(41) & 0.244(13) & 0.00e+00 & 6063(71) & 1.45(8) \\
159805154 & 14.80 &  0.90 & 2.686(9) & 0.29(34) & 5.95e-32 & 5642(50) & 1.50(23) \\
160583519 & 14.43 &  0.98 & 1.9731(7) & 0.175(11) & 0.00e+00 & 5994(52) & 1.292(48) \\
164754858 & 14.70 &  0.80 & 2.679(24) & 0.228(26) & 1.66e-11 & 5864(53) & 1.87(26) \\
165650748 & 11.38 &  1.00 & 2.11600(19) & 0.0768(18) & 0.00e+00 & 6626(123) & 1.401(26) \\
198240464 & 13.52 &  0.89 & 2.35724(7) & 0.816(6) & 0.00e+00 & 5646(53) & 1.195(23) \\
202466623 & 13.81 &  0.97 & 3.3207(7) & 0.237(11) & 0.00e+00 & 6009(60) & 1.155(30) \\
202507151 & 12.69 &  0.99 & 1.98289(36) & 0.135(5) & 0.00e+00 & 6356(69) & 1.032(27) \\
206688085 & 13.77 &  0.99 & 1.91975(28) & 2.337(9) & 0.00e+00 & 4936(47) & 2.43(7) \\
207208668 & 13.33 &  0.98 & 2.5671(6) & 0.36(36) & 0.00e+00 & 6111(52) & 1.166(37) \\
209397773 & 12.73 &  0.83 & 3.58(9) & 0.08(5) & 1.05e-69 & 5460(34) & 1.230(40) \\
212320065 & 11.64 &  0.98 & 1.8627(7) & 1.416(8) & 0.00e+00 & 5097(14) & 2.4992(9) \\
219988867 & 13.03 &  0.96 & 1.9209(19) & 0.1007(39) & 0.00e+00 & 6093(57) & 1.446(38) \\
228508601 & 14.40 &  0.99 & 5.58(6) & 0.346(24) & 0.00e+00 & 5400(41) & 1.97(33) \\
231845752 & 13.72 &  0.96 & 2.0923(35) & 0.088(8) & 1.51e-31 & 5864(62) & 1.69(5) \\
247017534 & 12.15 &  0.92 & 1.8154(46) & 0.09(10) & 0.00e+00 & 5945(28) & 1.355(49) \\
263014094 & 15.03 &  0.88 & 2.91052(14) & 1.531(25) & 0.00e+00 & 5000(9) & 1.64(8) \\
277892210 & 13.91 &  0.95 & 1.807791(24) & 1.658(28) & 0.00e+00 & 5460(49) & 1.234(32) \\
293463617 & 15.13 &  0.87 & 1.74504(12) & 1.337(40) & 0.00e+00 & 4612(39) & 0.802(39) \\
298093039 & 11.62 &  1.00 & 5.5610(6) & 0.1620(25) & 0.00e+00 & 5877(127) & 0.897(18) \\
304103779 & 12.96 &  0.99 & 1.12766(11) & 0.0599(47) & 9.54e-37 & 5968(29) & 1.542(36) \\
317129309 & 13.44 &  0.98 & 3.86(12) & 1.5(2.6) & 6.53e-24 & 6501(35) & 1.83(14) \\
320529836 & 14.65 &  0.88 & 0.8353(8) & 1.84(13) & 0.00e+00 & 5251(44) & 0.969(45) \\
320965274 & 12.66 &  0.99 & 1.759(6) & 0.0442(48) & 8.53e-22 & 5286(28) & 1.80(11) \\
325566833 & 13.35 &  0.98 & 1.0628(15) & 0.069(7) & 9.39e-23 & 5291(22) & 1.283(19) \\
335682563 & 13.55 &  0.99 & 4.4173(21) & 0.112(7) & 0.00e+00 & 5987(68) & 0.942(28) \\
346597868 & 11.52 &  0.97 & 1.782(6) & 0.0320(34) & 5.51e-28 & 6438(97) & 2.29(7) \\
349367583 & 13.05 &  0.99 & 2.302(14) & 0.23(24) & 0.00e+00 & 5919(53) & 1.465(49) \\
365771053 & 13.37 &  0.85 & 1.3515(6) & 1.089(14) & 0.00e+00 & 5981(27) & 1.139(20) \\
369371996 & 15.91 &  0.98 & 2.606(9) & 0.150(10) & 2.14e-40 & 5802(163) & 2.34(14) \\
382518318 & 14.01 &  0.29 & 4.33102(40) & 1.402(25) & 0.00e+00 & 5206(25) & 1.274(36) \\
389752750 & 15.24 &  0.82 & 0.67638(22) & 0.889(40) & 0.00e+00 & 4807(84) & 1.34(30) \\
393941149 & 14.73 &  0.86 & 3.873(29) & 0.364(25) & 0.00e+00 & 5119(45) & 1.04(11) \\
398940155 & 13.08 &  0.47 & 0.5039(7) & 0.62(12) & 0.00e+00 & 6976(65) & 2.10(14) \\
405471275 & 15.19 &  0.33 & 3.510(40) & 0.42(6) & 2.66e-12 & 5915(56) & 1.29(29) \\
406239686 & 13.45 &  0.98 & 1.633(5) & 0.12(13) & 1.16e-41 & 5499(32) & 1.395(46) \\
420049852 & 12.13 &  0.81 & 1.666(7) & 0.0538(46) & 6.95e-35 & 6313(58) & 1.361(42) \\
437237493 & 15.36 &  0.72 & 1.2648(39) & 0.84(42) & 0.00e+00 & 6186(60) & 1.41(19) \\
441401311 & 13.74 &  0.99 & 2.915(6) & 0.275(12) & 0.00e+00 & 5741(21) & 1.46(7) \\
461156754 & 14.96 &  0.74 & 1.610(6) & 0.321(42) & 2.32e-20 & 5730(59) & 1.57(17) \\
\end{longtable}
\end{center}

\begin{table*}[h!]
\caption{The 29 objects whose cause for variability is undetermined. Columns are the same as in Table~\ref{rot}.} 
\label{table:other}      
\centering          
\begin{tabular}{r r r r r r r r}
\hline\hline
TIC & $G$ & CROWDSAP & $P$ (days) & Amplitude (\%) & FAP & $T\eff^{\MS}$ (K) & $R^{\MS}$ ($R\sun$)\\
\hline
13069774 & 12.29 &  0.97 & 2.380(11) & 0.058(28) & 0.00e+00 & 5969(23) & 1.288(21) \\
31636688 & 16.01 &  0.78 & 2.806(15) & 0.39(6) & 2.88e-07 & 6974(36) & 1.62(37) \\
71133157 & 14.06 &  0.98 & 2.156(16) & 0.121(15) & 3.17e-10 & 6751(74) & 1.31(7) \\
80057233 & 15.79 &  0.62 & 6.53(12) & 2.98(43) & 1.49e-14 & 5986(73) & 1.78(38) \\
89529774 & 14.31 &  0.96 & 4.38(6) & 0.106(13) & 4.73e-08 & 5549(31) & 2.41(9) \\
141628019 & 14.34 &  0.95 & 10.42(5) & 0.060(9) & 2.75e-38 & 5518(28) & 2.32(18) \\
147115112 & 14.27 &  0.60 & 1.643(13) & 0.098(17) & 2.25e-04 & 6458(77) & 1.60(12) \\
152374958 & 13.08 &  0.73 & 4.86(6) & 0.093(11) & 2.04e-28 & 6015(23) & 2.36(9) \\
159669717 & 13.81 &  0.72 & 7.57(31) & 0.071(49) & 2.00e-07 & 6174(81) & 1.64(11) \\
181820016 & 12.82 &  0.99 & 10.26(44) & 0.055(8) & 4.67e-18 & 6518(129) & 1.48(8) \\
197693940 & 14.40 &  0.90 & 6.40(13) & 0.249(40) & 5.11e-12 & 4974(47) & 0.530(40) \\
220026025 & 15.67 &  0.87 & 9.85(7) & 18(10) & 2.26e-23 & 5346(63) & 1.8(7) \\
220472655 & 14.48 &  0.96 & 3.1301(46) & 0.078(10) & 6.23e-18 & 6734(93) & 1.35(6) \\
246881770 & 15.39 &  0.89 & 7.13(9) & 0.379(45) & 6.31e-07 & 5681(76) & 1.27(18) \\
253932935 & 16.00 &  0.89 & 13.4(1.8) & 0.29(47) & 3.86e-17 & 6159(110) & 1.88(40) \\
257024892 & 13.05 &  0.99 & 2.2077(16) & 0.0222(46) & 2.91e-07 & 6197(35) & 2.20(7) \\
259963278 & 10.61 &  1.00 & 0.83783(7) & 0.0089(8) & 1.15e-24 & 6495(20) & 1.460(24) \\
266347283 & 12.55 &  0.99 & 0.9011(11) & 0.0334(47) & 4.07e-15 & 6523(140) & 1.420(40) \\
274035031 & 13.96 &  0.46 & 2.825(13) & 0.090(16) & 2.41e-07 & 6055(22) & 1.666(49) \\
289737935 & 14.97 &  0.90 & 3.030(37) & 0.182(32) & 1.09e-04 & 5769(93) & 1.46(22) \\
313303167 & 12.46 &  0.88 & 0.4270633(17) & 0.7749(47) & 0.00e+00 & 7136(129) & 1.369(37) \\
320176500 & 15.79 &  0.46 & 6.39(38) & 0.65(26) & 6.34e-10 & 6158(48) & 1.88(38) \\
389520459 & 14.03 &  0.99 & 1.350(6) & 0.092(13) & 6.41e-06 & 5800(51) & 2.18(16) \\
410135274 & 15.41 &  0.25 & 6.96(8) & 0.86(6) & 7.30e-44 & 5497(22) & 2.04(25) \\
410390905 & 13.39 &  0.99 & 0.99130(10) & 0.061(5) & 2.77e-33 & 5900(45) & 1.576(26) \\
424941595 & 14.49 &  0.34 & 6.88(19) & 0.226(39) & 2.12e-15 & 5454(37) & 2.4991(12) \\
453366788 & 13.94 &  0.23 & 5.065(38) & 0.285(29) & 1.12e-33 & 6250(50) & 1.19(8) \\

\hline                  
\end{tabular}
\end{table*}

\begin{table*}
\caption{Systems for which no periodic variability has been detected. TIC, $G$, {\tt CROWDSAP}, $T\eff$, and radius are the same as in Table~\ref{rot}. The detection limit has been calculated as five times the average amplitude in a Fourier transform of the light curve.}
\label{table:NOV}      
\centering          
\begin{tabular}{r r r r r r}
\hline\hline       
TIC & $G$ & CROWDSAP & Detection limit (\%) & $T\eff^{\MS}$ (K) & $R^{\MS}$ ($R\sun$) \\ 
\hline                    
  9102069 & 15.04 &  0.88 & 0.17 & 6011(163) & 1.72(23) \\
 13090700 & 14.73 &  0.99 & 0.13 & 6689(57) & 1.85(17) \\
 25245570 & 15.12 &  0.69 & 0.28 & 5950(82) & 1.88(34) \\
 32661254 & 11.29 &  0.97 & 0.013 & 6425(53) & 1.74(7) \\
 56648314 & 14.13 &  0.99 & 0.09 & 5771(26) & 1.84(28) \\
 71150825 & 14.61 &  0.66 & 0.13 & 5382(17) & 1.35(8) \\
 71410075 & 15.87 &  0.75 & 0.30 & 6324(150) & 1.04(13) \\
 98871628 & 14.34 &  0.93 & 0.11 & 5697(24) & 1.13(7) \\
117626475 & 12.60 &  1.00 & 0.035 & 6161(112) & 1.05(9) \\
121550523 & 14.32 &  0.92 & 0.11 & 6128(69) & 1.49(7) \\
141602548 & 12.10 &  0.98 & 0.014 & 6706(171) & 1.700(36) \\
146437397 & 13.47 &  0.97 & 0.047 & 5797(59) & 1.63(5) \\
151892844 & 14.98 &  0.98 & 0.22 & 5047(69) & 0.57(8) \\
179278778 & 15.07 &  0.86 & 0.31 & 5855(96) & 2.04(24) \\
220370211 & 14.84 &  0.37 & 0.34 & 5782(70) & 1.26(10) \\
248949857 & 12.23 &  1.00 & 0.025 & 6507(28) & 1.82(14) \\
260839766 & 15.39 &  0.85 & 0.31 & 6065(213) & 1.89(25) \\
261427146 & 12.07 &  0.95 & 0.021 & 6070(121) & 1.7(7) \\
261679852 & 16.22 &  0.90 & 0.45 & 5095(64) & 1.47(34) \\
281851153 & 14.96 &  0.72 & 0.19 & 6501(35) & 1.83(14) \\
283866221 & 15.58 &  0.69 & 0.38 & 8117(218) & 1.38(42) \\
301405970 & 12.57 &  1.00 & 0.031 & 6606(87) & 2.03(10) \\
339525222 & 13.80 &  0.98 & 0.06 & 6017(26) & 1.87(8) \\
362105045 & 15.39 &  0.97 & 0.14 & 6400(79) & 1.05(10) \\
370282569 & 14.60 &  0.97 & 0.11 & 5845(54) & 2.22(15) \\
377053047 & 14.82 &  0.45 & 0.15 & 6550(50) & 1.46(15) \\
392703299 & 15.81 &  0.70 & 0.32 & 5431(109) & 2.13(23) \\
394631720 & 14.39 &  0.76 & 0.11 & 5311(58) & 1.29(7) \\
413300076 & 15.77 &  0.37 & 0.42 & 5117(74) & 1.7(6) \\
421951567 & 13.90 &  0.88 & 0.08 & 6216(46) & 2.08(21) \\
422149668 & 14.48 &  0.98 & 0.14 & 5608(62) & 1.71(13) \\
436639479 & 14.41 &  0.86 & 0.15 & 6297(83) & 0.773(49) \\
441399312 & 13.28 &  0.57 & 0.06 & 5499(124) & 1.89(10) \\
\hline                  
\end{tabular}
\end{table*}

\onecolumn
\begin{sidewaystable}
\caption{Identified common proper motion candidates. The four columns to the left refer to the hot subdwarf, whereas the next four refer to the candidate common proper motion companion. All values come directly from the {\it Gaia} DR2 catalogue. The last four columns contain pair properties, namely the angular separation $\theta$, the projected orbital separation $a$, assuming the distance of the hot subdwarf, and the differences in parallax and proper motion in units of standard deviation, $\Delta \varpi$ and $\Delta \mu$. The six objects marked with asterisks have identified close companions, being therefore candidate triple systems (see Section~\ref{cpm_cand}).}\label{cppm}
\centering
\begin{tabular}{ccccc|ccccc|cccc} 
\hline\hline
\multicolumn{5}{c|}{\it Hot subdwarf} & \multicolumn{5}{c|}{\it Candidate companion} & \multicolumn{4}{c}{\it Pair properties}\\
source\_id & $G$ & $\varpi$ & $\mu_{\alpha}$ & $\mu_{\delta}$ &
source\_id & $G$ & $\varpi$ & $\mu_{\alpha}$ & $\mu_{\delta}$ &
$\theta$ & $a$ & $\Delta \varpi$ & $\Delta \mu$\\
           & (mag) & (mas) & (mas/yr) & (mas/yr) &
           & (mag) & (mas) & (mas/yr) & (mas/yr) &
             ('') & (AU) &  & \\
\hline
  312628749626419328* & 13.1 & 1.73 &  -8.80 & -11.27 &   312628745331966976 & 18.7 & 1.47 &  -9.74 & -12.16 &  6.8 &  3920 & 1.1 & 2.8 \\
  601188910547673728 & 14.6 & 0.69 &  -5.02 &  -0.80 &   601188910547673600 & 15.4 & 0.58 &  -5.15 &  -0.75 &  2.7 &  3856 & 1.6 & 1.3 \\
  992534888766785024* & 12.0 & 2.93 & -25.48 & -25.16 &   992534888766784640 & 13.8 & 3.03 & -25.57 & -25.01 &  7.0 &  2384 & 1.3 & 1.3 \\
 1332156747638095488 & 15.9 & 0.46 &  -5.22 & -27.03 &  1332156747638095360 & 16.7 & 0.65 &  -5.41 & -27.29 &  4.1 &  8996 & 1.7 & 1.7 \\
 1659750327258228352 & 16.2 & 0.53 &  -1.74 &  -0.70 &  1659750327258228224 & 17.9 & 0.62 &  -2.02 &  -0.65 &  1.9 &  3595 & 0.7 & 1.3 \\
 1660055029417965952 & 13.4 & 2.18 & -18.82 &   3.51 &  1660055098137442944 & 17.5 & 2.10 & -18.82 &   3.95 & 39.8 & 18291 & 0.7 & 2.7 \\
 1883850072814402048 & 13.0 & 0.98 &  -1.80 & -13.31 &  1883850072814402432 & 17.4 & 0.96 &  -1.59 & -13.57 & 11.5 & 11655 & 0.2 & 1.9 \\
 1891098500140100352 & 12.8 & 2.11 &  11.43 &  -1.30 &  1891098500140101120 & 12.9 & 2.05 &  11.41 &  -1.17 & 26.0 & 12349 & 0.9 & 1.5 \\
 2002880555945732992 & 14.4 & 1.46 &   4.72 &   1.57 &  2002880555945731968 & 17.0 & 1.42 &   4.97 &   1.70 &  3.4 &  2346 & 0.4 & 2.2 \\
 2103959862471941632* & 14.1 & 1.55 &  -0.64 &  -8.95 &  2103959858173621760 & 18.8 & 1.50 &  -0.74 &  -8.72 & 15.9 & 10253 & 0.2 & 0.6 \\
 3381286602335612416 & 14.0 & 0.60 &  -3.11 &  -7.56 &  3381286636695992448 & 17.9 & 1.08 &  -2.70 &  -7.16 &  8.5 & 14220 & 2.9 & 2.0 \\
 3484319745326973824* & 11.3 & 1.95 &  -7.66 & -23.66 &  3484320501241217792 & 16.1 & 1.92 &  -8.10 & -23.65 &  6.4 &  3307 & 0.2 & 2.4 \\
 3868418219635118080* & 14.2 & 1.08 & -21.76 &  13.79 &  3868418219635275520 & 16.0 & 1.41 & -21.81 &  13.87 &  8.6 &  8019 & 1.9 & 0.2 \\
 4491274930955326080 & 15.6 & 0.61 &   0.08 & -10.03 &  4491274930955325440 & 16.8 & 0.49 &   0.27 & -10.06 &  4.9 &  8014 & 1.3 & 1.3 \\
 4877263019073081600* & 12.3 & 1.72 & 7.03 & 20.44 & 4877263023370516096 & 17.3 & 1.59 & 7.36 & 20.51 & 20.9 & 12144 & 1.6 & 2.5 \\
 1429755412672689536 & 13.7 & 1.30 &  -6.44 &  -4.82 &  1429755374017588608 & 12.6 & 1.43 &  -6.50 &  -4.50 &  3.6 &  2764 & 2.4 & 2.9 \\
 \hline
\end{tabular}
\end{sidewaystable}
\twocolumn

\begin{appendix}

\section{Two new $g$-mode pulsators}
\label{sec:puls}

The amplitude spectrum of TIC\,071013467 shows four significant frequencies with one just below our adopted threshold (S/N\,=\,4.5). Other sdBVs observed from space often show an asymptotic period spacing, which can be used for modal degree identification. Unluckily, the amplitude spectrum is too patchy and finding a complete sequence of consecutive overtones is very difficult. The period spacing between f$_4$ and f$_5$ is close to the expected 250\,sec, suggestive of dipole modes, however the spacings between the remaining frequencies do not indicate either single or multiples of around 250\,sec. They are also too large to identify quadrupole modes. We found no multiplet pattern, either. 

In the amplitude spectrum of TIC\,382518318 we detected two dominant frequencies in the $g$-mode region, and one low amplitude frequency, too short to be considered a $g$-mode. All these frequencies meet our S/N\,=\,4.5 criterion, where N\,=\,0.035\,ppt, though suggestive frequencies below this threshold are also considered. In addition, we detected one frequency, f$_3$ with S/N\,=\,4.1 close to f$_2$, along with two other frequencies, between f$_2$ and f$_3$, but with even smaller amplitudes falling way below our threshold. All these four frequencies suggest a quintuplet with one component missing, and an average frequency splitting around 1.45\,$\mu$Hz, which translates to a rotation period of 4\,days. This would be the shortest period in sdB stars detected thus far. On the other hand, if a smaller amplitude frequency shows rotational splitting, we could expect the higher amplitude frequency would show such a signature even clearer, since $g$-modes cannot be {\it l}\,=\,0. The highest amplitude frequency shows no multiplet splitting. We can always invoke that the split components are not driven but this interpretation is less likely.
The low frequency f$_1$ can be interpreted by either a binary signature or a contamination by neighbouring stars. The latter can easily be excluded by no presence of bright stars that could contaminate the target flux in large TESS pixels (21\,arc\,sec on side). We therefore conclude that f$_1$ is likely the signature of a binarity.

\begin{table}[h!]
\caption{Solutions for the two newly identified pulsators.}
\label{table:puls}
\centering 
\begin{tabular}{c c c c c}
\hline\hline
\multicolumn{5}{c}{{TIC~071013467 ($G = 13.4$)}} \\
Mode        & \multicolumn{2}{c}{{Frequency}} & Amplitude & S/N\\
            &  ($\mu$Hz)     &  (sec)    &  (ppt)  & \\
\hline
f$_{\rm 1}$ &  250.105(38) &   3998.3(6) &   0.50(8) &   5.3\\
f$_{\rm 2}$ &  320.392(38) &   3121.17(37) &   0.50(8) &   5.3\\
f$_{\rm 3}$ &  328.782(44) &   3041.53(41) &   0.43(8) &   4.6\\
f$_{\rm 4}$ &  400.954(25) &   2494.05(16) &   0.76(8) &   8.1\\
f$_{\rm 5}$ &  443.929(45) &   2252.61(23) &   0.42(8) &   4.4\\
\hline\hline
\multicolumn{5}{c}{{TIC~158235404 ($G = 11.6$)}} \\
Mode        & \multicolumn{2}{c}{{Frequency}} & Amplitude & S/N\\
            &  ($\mu$Hz)     &  (sec)    &  (ppt)  & \\
\hline
f$_{\rm 1}$ &   17.801(36) &   56178(112) &   0.188(30) &   5.3\\
f$_{\rm 2}$ &  184.484(17) &   5420.52(49) &   0.397(30) &  11.2\\
f$_{\rm 3}$ &  190.132(45) &   5259.5(1.3) &   0.147(30) &   4.1\\
f$_{\rm 4}$ &  232.006(6) &   4310.23(11) &   1.126(30) &  31.8\\
\hline  
\end{tabular}
\end{table}

\onecolumn
\section{ADQL query for co-moving sources within 20\,000~AU}
\label{adql}

\begin{verbatim}
SELECT sd.NAME, sd.source_id, sd.ra, sd.dec, sd.parallax, sd.parallax_error,
       sd.pmra, sd.pmdec, sd.pmra_error, sd.pmdec_error,
       g.source_id, g.ra, g.dec, g.parallax, g.parallax_error,
       g.pmra, g.pmra_error, g.pmdec, g.pmdec_error,
       g.phot_g_mean_mag, g.bp_rp, g.phot_bp_rp_excess_factor,
       g.astrometric_chi2_al, g.astrometric_n_good_obs_al,
       DISTANCE(POINT('ICRS', sd.ra, sd.dec), POINT('ICRS', g.ra, g.dec)) AS dist
FROM [INPUT TABLE] AS sd
JOIN gaiadr2.gaia_source AS g
ON 1=CONTAINS(POINT('ICRS', sd.ra, sd.dec), CIRCLE('ICRS', g.ra, g.dec, 20.0*sd.parallax/3600.))
WHERE (sd.source_id != g.source_id)
AND ( ( (sd.pmra - g.pmra)*(sd.pmra - g.pmra)/
(sd.pmra_error*sd.pmra_error + g.pmra_error*g.pmra_error) +
(sd.pmdec - g.pmdec)*(sd.pmdec - g.pmdec)/
(sd.pmdec_error*sd.pmdec_error + g.pmdec_error*g.pmdec_error) ) <= 9.0 )
AND ( (sd.parallax - g.parallax)*(sd.parallax - g.parallax)/
(sd.parallax_error*sd.parallax_error + g.parallax_error*g.parallax_error) <= 9.0 )
\end{verbatim}
\twocolumn

\section{Hot subdwarfs with candidate common proper motion companions}
\label{cpm_cand}

The 16 objects in the hot subdwarf catalogue of \citet{geier2020} with identified possible common proper motion companions are described in more detail below. We refer to the objects using the {\it Gaia} DR2 source id, but also include their identification as in \citet{geier2020} for readability.

{\bf 312628749626419328 (GALEX~J01012+3125):} it is classified as sdB+MS in \citet{geier2020}. The available LAMOST spectra show instead a single sdB, and the {\it Gaia} $G_{BP} - G_{RP}$ colour of $-0.37$ also suggests there is no composite companion. The classification might rely on photometry that could have been contaminated by the common proper motion companion, which is 6.8" away. There is, on the other hand, a shift of $37\pm2$~km/s between LAMOST spectra taken 678~days apart, which cannot be explained by the common proper motion companion, suggesting that there is an inner binary and that this could be a triple system.

{\bf 601188910547673728 (LAMOST~J082517.99+113106.3):} also classified as sdB+MS \citep{geier2020}, but in this case the classification would seem to be confirmed at first glance by the available LAMOST spectrum. However, the size of the LAMOST fibres of 3.3" means that the common proper motion companion, which is only 2.7" away and has similar brightness, contaminates the spectrum.

{\bf 992534888766785024 (GALEX~J063952.00+515658.00):} it is classified as a single sdB \citep{geier2020}, but \citet{nemeth2012} suggest it could be a close binary because of the large radial velocity ($> 100$~km/s) with respect to the kinematic local standard of rest (LSR). This is thus another candidate to a triple system.

{\bf 1332156747638095488 (PG~1623+386):} classified as sdOB \citep{geier2020}, it has been flagged as a visual double by \citet{saffer1998}. This is likely because of the common proper motion pair at a distance of 4.1".

{\bf 1883850072814402048 (GALEX~J22484+2714):} the literature on this object is scarce; it has only been included in the catalogues of \citet{geier2019} and \citet{geier2020} as sdB+MS. The brightness ($G = 13.4$) and relative red colour ($G_{BP} - G_{RP}$ = 0.26) suggest it might be instead a misclassified blue horizontal branch (BHB) star. We note that the companion is almost 40" away, and therefore should not contaminate the {\it Gaia} colour.

{\bf 2103959862471941632 (Kepler~J19028+4134):} this star has been observed by the Kepler mission. Whereas \citet{ostensen2010} reported no pulsations compatible with a hot subdwarf, \citet{mcnamara2012} identified a peak at 0.21~cycles/day with an amplitude of 0.0194\%, and classified the object as a possible slowly rotating B star. However, {\it Gaia} places this object within the hot subdwarf cloud ($M_G = 9.7, G_{BP} - G_{RP} = -0.46$), which suggests that the variability has instead another origin, likely related to a binary companion other than the common proper motion pair, since such period is not expected to be observed in a single hot subdwarf or due to a distant ($\theta$ = 16") companion, making this system another possible triple.

{\bf 3484319745326973824 (EC~11429-2701):} this object has also been observed by TESS (TIC~32661254) and was included in our analysis of Section~\ref{TESS}. It was observed and classified as part of the Edinburgh-Cape Blue Object Survey \citep[EC Survey,][]{kilkenny1997}, and our SED fit indicates the presence of a composite companion with $T\eff \simeq 6400$~K. The common proper motion companion is 6.44" away, is five magnitudes fainter than the hot subdwarf, and has its own photometry measurements in VizieR, therefore it should not have affected our SED fit. This is thus likely a triple system, with an inner unresolved binary.

{\bf 3868418219635118080 (GALEX~J11009+1055):} it has been classified as a composite hot subdwarf by \citet{nemeth2012}, with a F6V type companion showing $T\eff \approx 6430$~K. Given that the common proper motion companion is resolved and more than 8.5" away, it is rather unlikely that it contaminated the spectrum. Moreover, the {\it Gaia} DR2 archive suggests a $T\eff = 4750^{+245}_{-128}$~K and $R = 0.578^{+0.032}_{-0.027}~R\sun$ for the common proper motion companion, which is more compatible with a late-G or early-K main sequence star. Therefore this is possibly a hierarchical triple system.

{\bf 4877263019073081600 (EC~05015-2831):} it is a known composite system that has also been included in the analysis of Section~\ref{TESS} (TIC~13069774, shown in Fig.~\ref{othervar}). The common proper motion companion is at almost 21" away, having no contribution to the SED. This is another triple candidate.

{\bf 1429755412672689536 (PG~1618+562):} although it is classified as sdBV+F3 in \citet{geier2020}, \citet{drilling2013} include this object as a single sdB reference star. The colour-excess which lead to the composite classification is actually due to the common proper motion companion at 3.6", which was previously identified by \citet{silvotti2000}.

For the remaining systems (1659750327258228352 = PG 1411+590; 1660055029417965952 = GALEX J14085+5940; 1891098500140100352 = FBS 2253+335; 2002880555945732992 = KPD 2254+5444; 3381286602335612416 = LAMOST J065446.63+244926.8; 4491274930955326080 = SDSS J172125.76+090311.2), the literature refers mostly to their inclusion in hot subdwarf catalogues. There is no study on their binarity to the best of our knowledge.

\end{appendix}

\end{document}